\documentclass[aps,prx,reprint,amsfonts,amsmath,amssymb,longbibliography]{revtex4-2}
\usepackage[usenames]{color}
\usepackage{hyperref}
\usepackage{graphicx}
\usepackage{bm}
\usepackage{newtxtext}
\usepackage[varg]{newtxmath}
\usepackage[normalem]{ulem}

\hypersetup{colorlinks=true,linkcolor=blue,citecolor=blue,urlcolor=blue}

\begin{document}
\title{Energy-band echoes:\\ Time-reversed light emission from optically driven quasiparticle wavepackets}
\author{Shohei Imai}
\author{Atsushi Ono}
\author{Sumio Ishihara}
\thanks{Deceased.}
\affiliation{Department of Physics, Tohoku University, Sendai 980-8578, Japan}
\date{\today}

\begin{abstract}
The at-will control of quantum states is a primary goal of quantum science and technology.
The celebrated Hahn echo exemplifies such quantum-state control based on a time-reversal process in a few-level system.
Here, we propose a different echo phenomenon associated with the energy-band structure in quantum many-body systems.
We show that the dynamics of quasiparticle wavepackets can be reversed by a driving electric-field pulse, yielding echoes with the time-reversed waveform of the optical excitation pulse when the quasiparticles recombine.
The present echoes are observed not only in band insulators but also in correlated insulators, including a Mott insulator and a spontaneously-broken-symmetry charge-ordered insulator, in one- and higher-dimensional systems, irrespective of the integrability of the models.
Analytical expressions reveal the conditions under which the echoes appear, and they also indicate that the frequency of the echo pulses reflects the dispersion relation for quasiparticles such as electron--hole pairs, doublon--holon pairs, and kink--antikink pairs.
These findings provide a framework for all-optical momentum-resolved spectroscopy of the quasiparticles in quantum many-body systems.
\end{abstract}
\maketitle

%================================================================
%================================================================
\section{Introduction}
Precise control of quantum systems is of great importance for the development of quantum science and technology~\cite{Kjaergaard2020,Preskill2018,Bruzewicz2019,Kok2007,Awschalom2018,Xu2020,Degen2017,Pirandola2018,Atature2018,OBrien2009,Brif2010,Brecht2015}.
Well-known examples of quantum control include the Hahn echoes---also called spin echoes---in few-level quantum systems~\cite{Hahn1950}.
This phenomenon is the time-reversal refocusing of quantum spins, and it has been widely used in many areas, since they provide access to the spectroscopic properties and relaxation dynamics of quantum systems via nuclear magnetic resonance.
Such a time-reversal process is a manifestation of unitary evolution and thus of the controllability of the quantum systems; and it can also be observed as photon echoes in condensed matter systems~\cite{Abella1966}, Loschmidt echoes in quantum physics~\cite{Gorin2006,Goussev:2012,Sanchez2016}, and as what is often called a ``time mirror'' in photonic, phononic, and fermionic systems~\cite{Yanik2004,Longhi2007a,Sivan2011a,Wimmer2018,Chumak2010,Yuan2016,Reck2017}.

Coherent optical control of quantum many-body systems has also been recognized as an intriguing problem in modern condensed matter physics, since such a system exhibits a wide variety of phases and thereby provides a rich playground for the control of physical properties.
Recent developments in photoinduced phase transitions~\cite{Koshihara2022,Ishihara2019,Miyamoto2018,Kirilyuk2010} and Floquet engineering~\cite{DelaTorre2021,Rudner2020,Harper2020,Oka2018,Mentink2017} have attracted growing interest in studying the ultrafast and nonthermal control of quantum materials, owing to the rapid development of light sources~\cite{Fulop2020,Kim2021,Krausz2009,Goulielmakis2007} and of time-resolved measurement techniques~
\cite{Koshihara2022,DelaTorre2021,Mandal2021,Suzuki2021a,Smallwood2016,Kheifets2020,Wen2019,Reid2016,Sidiropoulos2021,McIver2020,Gerber2017,Wang2013,Wen2019,Danz2021,RubianodaSilva2018,Reid2016,Kheifets2020}.
However, the coherent control of condensed matter systems still faces challenges, since quantum coherence is quickly destroyed by interactions with environmental degrees of freedom.

Another strategy for ultrafast control of many-body dynamics is to use high-harmonic generation (HHG), which is the nonperturbative process of attosecond pulse generation from electrons driven coherently by a lightwave.
Originally, HHG was investigated in atomic gases~\cite{McPherson1987,Ferray1988,Kfir2015,Gaumnitz2017,Agostini2004,Gallmann2012,Li2020c,Lewenstein2021}, and it has recently been studied in semiconductors~\cite{Ghimire2011,Schubert2014a,Hohenleutner2015,Garg2016a,Langer2016,Langer2017,Huttner2017,Kruchinin2018,Ortmann2021,Xia2021,Tamaya2016}, topological materials~\cite{Yoshikawa2017,Bai2020a,Schmid2021,Lv2021}, and strongly correlated systems~\cite{Silva2018,Murakami2018d,Tancogne-Dejean2018a,Nag2019,Lysne2020a,Granas,Imai2019q,Fauseweh2020b,Murakami2021,Bionta2021,Uchida2021b,Robson2017,Yang2019a,Alcala2022}.
The HHG mechanism can basically be understood from the semiclassical theory called the three-step model~\cite{Corkum1993,Lewenstein1994,Vampa2015}: the electrons are ionized (or excited to conduction bands), accelerated, and then recombined during a single optical cycle of an external pulse.
This process involves the coherent reciprocating motion of the electrons controlled by the external pulse; we therefore anticipate that the real-time HHG profile can be understood in terms of the time-reversal process in such a many-body system.
Furthermore, since the electrons are adiabatically accelerated within energy bands in crystalline solids, it has been recognized that HHG spectra contain information about the energy-band structure~\cite{Luu2015a,Vampa2015b,Wang2016a,Garg2016a,You2017c,Stepanov2017b,Yu2018c,Kaneshima2018,Li2019e,Uzan2020a,Lakhotia2020,Chen2021f}, including the Berry curvature~\cite{Liu2017b,Banks2017,Luu2018,Avetissian2020b,Lou2021} and the momentum-dependent transition dipole moments~\cite{Zhao2019,Lu2019j,Uchida2021a}.
However, these methods require numerical simulations consistent with the experiments, and they are hard to apply to strongly correlated systems.

In this paper, we investigate the real-time dynamics of quasiparticle wavepackets driven by a lightwave.
We discover an echo phenomenon associated with the energy-band structure in crystalline solids, which we term ``energy-band echoes.''
We first demonstrate the appearance of the echoes of an optical excitation process in a minimal one-dimensional model of a band insulator, performing numerically exact simulations of the real-time evolution, and we then derive analytical expressions for the echoes.
We find that the echoes are generated after photoexcited quasiparticles are driven by a half-cycle pulse that reverses the group velocity of the wavepackets and thereby achieves a time-reversal process, i.e., the recombination of the wavepackets.
We also find that the dispersion relation of the electron--hole pair is reflected in the frequency of the echo as a function of the pulse amplitude.
Furthermore, we confirm that the echoes emerge from quasiparticles in correlated insulators described by the Hubbard model and by the transverse-field Ising model, and we show that the echo frequency is consistent with the predictions obtained from the exact solutions.
The echoes are observed even in a two-dimensional system and a non-integrable system, implying the generality and applicability of such energy-band echoes.
These findings suggest that the energy-band echoes make possible all-optical reconstructions of well-defined dispersion relations not only for the electron--hole pair but also for renormalized quasiparticles such as doublon--holon pairs and kink--antikink (domain wall) pairs.

The rest of this paper is organized as follows.
In Sec.~\ref{sec:echo}, we introduce the concept of energy-band echoes on the basis of numerical simulations and analytical expressions.
In Sec.~\ref{sec:spectroscopy}, we show that the echo signal reflects the energy and momentum of photoexcited quasiparticles in several types of insulators and that these echoes can be applied to all-optical momentum-resolved spectroscopy.
Sections~\ref{sec:discussion} and \ref{sec:summary}, respectively, are devoted to a discussion and a summary of this work.

%================================================================
%================================================================
\section{Energy-band echoes} \label{sec:echo}
In this section, we discuss the dynamics of photoexcited carriers that are induced in band insulators by two pulses with different frequencies.
The first pulse is a weak optical pulse that creates electron--hole pairs inside an energy band; we assume that its spectral width is much narrower than the energy-band width.
The subsequent second pulse is an off-resonant half-cycle pulse that adiabatically drives the electrons but does not induce interband transitions.
Hereafter, we term the first and second pulses the excitation pulse and the driving pulse, respectively.
The vector potential of the light field can be written as $\bm{A}(\tau) = \bm{A}_{\rm e}(\tau) + \bm{A}_{\rm d}(\tau)$, where $\bm{A}_{\rm e}(\tau)$ and $\bm{A}_{\rm d}(\tau)$ are the vector potentials of the excitation pulse and the driving pulse, respectively, at time $\tau$.
The electric field is given by $\bm{E}(\tau) = -\partial_{\tau} \bm{A}(\tau)$.
Throughout this paper, the Dirac constant, electron charge, and lattice constant are set to unity.

\subsection{Numerical simulation} \label{sec:asym_echo}
First, we demonstrate the appearance of echoes in a band insulator by simulating the real-time dynamics induced by the excitation pulse and driving pulse.
We consider a tight-binding model of a two-orbital band insulator.
The Hamiltonian is given by
\begin{align}
\mathcal{H}_{\rm bi}=&-\sum_{\langle i,j \rangle} \sum_{\nu\nu'} t^{\nu\nu'}_{ij} c_{\nu i}^{\dagger} c_{\nu' j} 
+\sum_{i \nu} D_{\nu} c_{\nu i}^{\dagger} c_{\nu i}, 
\label{eq:biH}
\end{align}
where $c_{\nu i}^{\dagger}$ ($c_{\nu i}$) is the creation (annihilation) operator for an electron in orbital $\nu\ (=\alpha,\beta)$ at site $i$.
The first term in Eq.~\eqref{eq:biH} represents nearest-neighbor electron hopping, with $t^{\nu \nu'}_{ij}$ being the transfer integral.
The intraorbital and interorbital transfer integrals are set to $t^{\alpha\alpha}_{ij} = t^{\beta\beta}_{ij} = t_{\rm h}$ and $t^{\alpha\beta}_{ij} = t^{\beta\alpha}_{ij} = t^{\alpha\beta}$, respectively.
The vector potential $A(\tau)$ is introduced via the Peierls substitution: $t_{\rm h} \to t_{\rm h} \exp[-\mathrm{i} \bm{A}(\tau) {\cdot} (\bm{r}_j - \bm{r}_i)]$ and $t^{\alpha\beta} \to t^{\alpha\beta} \exp[-\mathrm{i}\bm{A}(\tau) {\cdot} (\bm{r}_j - \bm{r}_i)]$, with $\bm{r}_i$ being the position of site~$i$.
The second term in Eq.~\eqref{eq:biH} describes the on-site energy of each orbital; we assume $D_{\alpha}=-D_{\beta}=E_{\rm g}/2$, with $E_{\rm g}$ being the energy gap.
The interorbital transfer integral and the energy gap are set to $t^{\alpha\beta} = 2t_{\rm h}$ and $E_{\rm g} = 3t_{\rm h}$.
Energy and time are expressed in units of $t_{\rm h}$ and $t_{\rm h}^{-1}$, respectively.

%===============================================
\begin{figure}[t]
\centering
\includegraphics[width=\columnwidth]{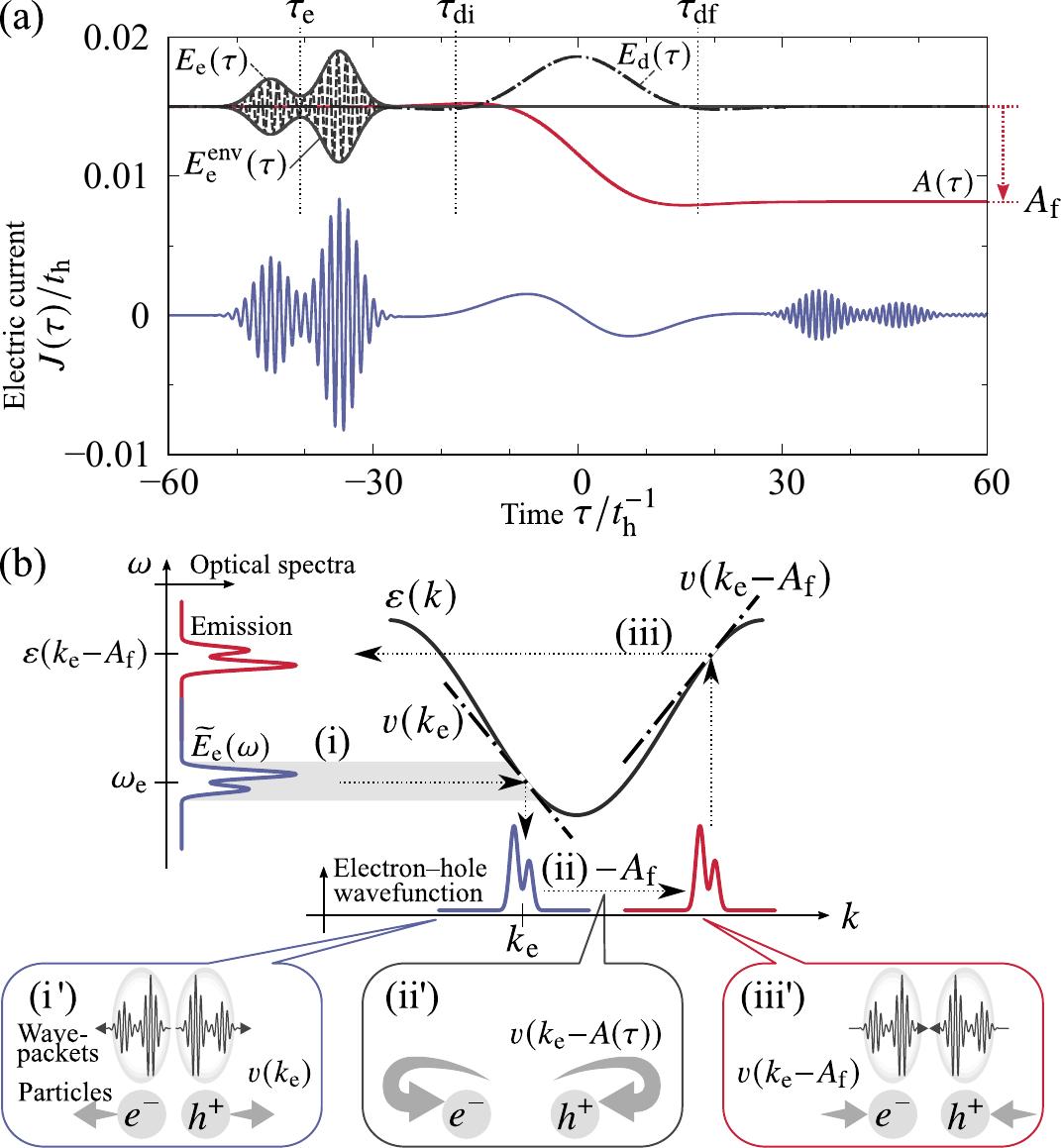}
\caption{(a)~Time profile of the electric current in a one-dimensional band insulator.
The dashed and dashed--dotted curves represent the electric fields of the excitation pulse $E_{\rm e}$ and the driving pulse $E_{\rm d}$, respectively.
The parameter values are set to $\omega_{\rm e} = 4\pi\sigma_{\rm e}^{-1} =5t_{\rm h}$, $\omega_{\rm d}=\sigma_{\rm d}^{-1}=0.1t_{\rm h}$, $A_{\rm e}=0.002$, $A_{\rm d}=0.9$, and $N=1500$.
(b)~Sketch of the echo-generation process, which consists of the following three steps:
(i)~creation of a photocarrier by the excitation pulse,
(ii)~intraband acceleration by the driving pulse,
and (iii)~echo emission due to the recombination of the photocarrier wavepacket.
See text for details.
}
\label{fig:echo_bi}
\end{figure}
%===============================================

In this section, we consider a one-dimensional chain and impose the periodic boundary condition.
The numbers of sites and electrons are denoted by $L$ and $N$, respectively, and the electron density is set to $N/L=1$.
The energy of a photoexcited electron--hole pair is given by
\begin{align}
\varepsilon_{\rm 1bi}(k) = 2\sqrt{(2t^{\alpha\beta}\cos k )^2+(E_{\rm g}/2)^2},
\label{eq:1bi_disp} 
\end{align}
where $k$ is the momentum of the electron.
We simulate the time evolution numerically by using the equation $|\psi(\tau+\delta\tau)\rangle = \exp [-\mathrm{i}\mathcal{H}(\tau+\delta\tau/2) \delta\tau ] |\psi(\tau)\rangle + \mathcal{O}(\delta\tau^3)$, with $\delta \tau = 0.01t_{\rm h}^{-1}$, and we calculate the electric current $J(\tau) = \langle \psi(\tau) \vert \hat{J}(\tau) \vert \psi(\tau) \rangle$, where $\hat{J}(\tau)=-N^{-1} \delta \mathcal{H}/\delta A(\tau)$.
The excitation pulse is given by
\begin{align}
A_{\rm e}(\tau) &= - \left[ \frac{1}{2} \mathrm{e}^{-(\tau-\tau_{\rm e}+2\sigma_{\rm e})^2/(2\sigma_{\rm e}^2)} + \mathrm{e}^{-(\tau-\tau_{\rm e}-2\sigma_{\rm e})^2/(2\sigma_{\rm e}^2)} \right] \notag \\
&\quad \times A_{\rm e} \sin(\omega_{\rm e}\tau), \label{eq:excitationpulse_sec2}
\end{align}
where $A_{\rm e}$ and $\omega_{\rm e}$ denote the amplitude and frequency of the pulse, respectively, and $\sigma_{\rm e}$ represents the pulse width.
Here, the envelope function is chosen so that the time-reversed waveform can be identified easily, as shown in Fig.~\ref{fig:echo_bi}(a).
The electric field of the driving pulse is given by
\begin{align}
E_{\rm d}(\tau) = E_{\rm d} \mathrm{e}^{-\tau^2/(2\sigma_{\rm d})^2} \cos(\omega_{\rm d} \tau + \theta_{\rm d}), \label{eq:drivingpulse}
\end{align}
where $E_{\rm d}$ ($= A_{\rm d}\omega_{\rm d}$), $\omega_{\rm d}$, $\sigma_{\rm d}$, and $\theta_{\rm d}$ denote the amplitude, frequency, pulse width, and carrier envelope phase (CEP), respectively.
Since the driving pulse is preferably an off-resonant half-cycle pulse, these parameters should satisfy $\omega_{\rm d} \ll E_{\rm g}$ and $\omega_{\rm d} \sigma_{\rm d} \sim 1$.
Note that the vector potential of the driving pulse remains finite after irradiation, and its value is given by
\begin{align}
A_{\rm f} = -\int_{-\infty}^{\infty} \mathrm{d}\tau\, E_{\rm d}(\tau) = -A_{\rm d} \sqrt{2\pi} \omega_{\rm d} \sigma_{\rm d} \mathrm{e}^{-(\omega_{\rm d}\sigma_{\rm d})^2/2} \cos\theta_{\rm d}.
\end{align}
Hereafter, we discuss the energy-band echo with $\theta_{\rm d}=0$, which facilitates the analysis.
We note that the existence of such a unipolar pulse with $A_{\rm f} \neq 0$ is controversial~\cite{Bessonov1981,Kim2000b,Arkhipov2020,Arkhipov2022}; we demonstrate in Sec.~\ref{sec:optical} that the echoes can be generated even by a driving pulse with $\theta_{\rm d} = \pi/2$; i.e., $A_{\rm f} = 0$.

Figure~\ref{fig:echo_bi}(a) shows the calculated electric current induced by the excitation pulse and by the driving pulse.
The delay time between the two pulses is set to $-\tau_{\rm e} = 40 t_{\rm h}^{-1}$.
When the excitation pulse arrives, at $\tau \approx \tau_{\rm e}$, the electric current oscillates in proportion to $E_{\rm e}(\tau)$.
We find that, following the driving pulse, the electric current oscillates again, even though an electric field is absent, and the envelope of this oscillation is a precisely time-reversed waveform of the excitation pulse.
Since the electromagnetic radiation is proportional to $\partial_{\tau} J(\tau)$, this oscillation of the electric current emits light with the time-reversed waveform, which originates from the quasiparticles in energy bands, as discussed in Sec.~\ref{sec:analysis} below.
We therefore term this phenomenon the ``energy-band echo'' in this paper.

\subsection{Analytical calculation} \label{sec:analysis}
In this section, we derive analytical expressions for the echo signal and elucidate the mechanism of echo generation, focusing again on a one-dimensional system for simplicity.
We consider a two-orbital band insulator, as in Sec.~\ref{sec:asym_echo}, but we introduce the electric-field pulses as electric-dipole couplings, not as the Peierls substitution of the vector potentials.
The Hamiltonian is written as $\mathcal{H}_{\rm 2band} = \mathcal{H}_0 + \mathcal{H}_{\rm e} + \mathcal{H}_{\rm d}$, where
\begin{align}
\mathcal{H}_0 &= \sum_{k} \sum_{\lambda\in \{ \rm cb,vb \}} \varepsilon^{\lambda}(k) c_{\lambda,k}^{\dagger} c_{\lambda,k}, \label{eq:2bandH_0} \\
\mathcal{H}_{\rm e}(\tau)&= -\sum_k E_{\rm e}(\tau) { \left( \mu_k c_{{\rm cb},k}^{\dagger} c_{{\rm vb},k}  +\mu_k^{*} c_{{\rm vb},k}^{\dagger} c_{{\rm cb},k} \right) }, \label{eq:2bandH_e} \\
\mathcal{H}_{\rm d}(\tau)&= -\sum_{j\lambda} E_{\rm d}(\tau) r_j c_{\lambda,j}^{\dagger} c_{\lambda,j}. \label{eq:2bandH_d}
\end{align}
Here, $c_{\lambda,k}^{\dagger}$ $(c_{\lambda,k})$ is the creation (annihilation) operator for an electron with momentum $k$ in orbital $\lambda \in \{ \mathrm{cb}, \mathrm{vb} \}$, where $\mathrm{cb}$ and $\mathrm{vb}$ denote the conduction and valence bands, respectively.
The energy of an electron--hole pair is given by $\varepsilon(k) = \varepsilon^{\rm cb}(k)-\varepsilon^{\rm vb}(k)$.
The term $\mathcal{H}_0$ in Eq.~\eqref{eq:2bandH_0} describes the non-interacting electrons.
The light-matter interactions are introduced through the time-dependent interband and intraband dipole Hamiltonians, $\mathcal{H}_{\rm e}(\tau)$ and $\mathcal{H}_{\rm d}(\tau)$, respectively.
Here, $\mu_{k}$ is the transition dipole moment, and $c_{\lambda,j}^{\dagger}$ is defined by $c_{\lambda,j}^{\dagger} = N^{-1/2} \sum_k \mathrm{e}^{-\mathrm{i}kr_j} c_{\lambda,k}^{\dagger}$.
Assuming that the excitation pulse is weak and that the driving pulse is off-resonant, we omit the negligible contributions of $E_{\rm e}$ to $\mathcal{H}_{\rm d}$ and of $E_{\rm d}$ to $\mathcal{H}_{\rm e}$~\cite{Meier1994}.
Since the echoes appear after the driving pulse, we consider the interband polarization and electric current defined by
\begin{align}
\hat{P}_{\mathrm{inter}} =- \frac{1}{N} \frac{\delta\mathcal{H}_{\rm e}}{\delta E_{\rm e}} = \frac{1}{N} \sum_k \left( \mu_k^{*} c_{{\rm vb},k}^{\dagger} c_{{\rm cb},k} + \mathrm{H.c.} \right)
\end{align}
and
\begin{align}
\hat{J}_{\mathrm{inter}} &= -\mathrm{i}[\hat{P}_{\mathrm{inter}},\mathcal{H}_0+
\mathcal{H}_{\mathrm{e}}] \notag \\
&= \frac{1}{N} \sum_k \varepsilon(k) { \left( -\mathrm{i} \mu_k^{*} c_{{\rm vb},k}^{\dagger} c_{{\rm cb},k} + \mathrm{H.c.} \right) }, \label{eq:analytical_currentop}
\end{align}
respectively.
The many-body state at time $\tau$ can be written as
\begin{align}
\vert\psi(\tau)\rangle = \prod_k {\left[ \psi^{\rm cb}_k(\tau) c_{{\rm cb},k}^{\dagger} + \psi^{\rm vb}_k(\tau) c_{{\rm vb},k}^{\dagger} \right]} \vert 0 \rangle, \label{eq:analytical_state}
\end{align}
where $\psi^{\rm cb}_k$ and $\psi^{\rm vb}_k$ are the probability amplitudes of the conduction- and valence-band states with momentum $k$ that satisfy the relation $|\psi^{\rm cb}_k(\tau)|^2+|\psi^{\rm vb}_k(\tau)|^2=1$, and $\vert 0 \rangle$ is the vacuum.
By using Eqs.~\eqref{eq:analytical_currentop} and \eqref{eq:analytical_state}, we obtain the expectation value of the interband current as
\begin{align}
J_{\mathrm{inter}}(\tau) = \frac{1}{N} \sum_k \varepsilon(k) { \left[ -\mathrm{i} \mu_k^{*} \psi^{\rm vb}_k(\tau)^{*} \psi^{\rm cb}_k(\tau) + \mathrm{c.c.} \right] }.
\label{eq:2band_current}
\end{align}

We next summarize the setup of the external fields and the conditions that must be met.
The excitation pulse is given by
\begin{align}
E_{\rm e}(\tau) = \frac{1}{2} E_{\rm e}^{\rm env}(\tau) \mathrm{e}^{-\mathrm{i}(\omega_{\rm e}\tau+\theta)} + {\rm c.c.},
\end{align}
where $\omega_{\rm e}$ denotes the carrier frequency, $\theta$ represents a phase constant, and $E_{\rm e}^{\rm env}(\tau)$ is a slowly varying envelope function centered at time $\tau = \tau_{\rm e}$~\footnote{$E_{\rm e}^{\rm env}(\tau)$ can be a complex function for, e.g., a chirped excitation pulse.
If this is the case, however, the present analysis remains valid.}.
The Fourier spectrum of the excitation pulse is given by $\widetilde{E}_{\rm e}(\omega) = \int_{-\infty}^{\infty} \mathrm{d}\tau\, E_{\rm e}(\tau) \mathrm{e}^{\mathrm{i}\omega\tau}$.
The spectral width of $\widetilde{E}_{\rm e}(\omega)$ must be narrow, and it must be located in the continuum of the interband excitation; these conditions can be written as $\int_{0}^{\infty} \mathrm{d}\omega\, (\omega {-} \omega_{\rm e})^2 |\widetilde{E}_{\rm e}(\omega)| {\big/} \int_{0}^{\infty}\mathrm{d}\omega\, |\widetilde{E}_{\rm e}(\omega)| \ll W^2$ and $E_{\rm g} < \omega_{\rm e} < E_{\rm g} + W$, where $E_{\rm g}$ and $W$ are the energy gap and the bandwidth of the photocarriers, respectively.
The peak amplitude of $E_{\rm e}(\tau)$ is assumed to be small enough so that the excited state is in the linear-response regime.
After the excitation pulse decays, the driving pulse $E_{\rm d}(\tau)$ is applied from $\tau=\tau_{\rm di} \ (\gg \tau_{\rm e})$ to $\tau=\tau_{\rm df}$.
The central frequency of the driving pulse is much less than the bandgap $E_{\rm g}$, which prevents unwanted interband excitations.
After irradiation by the driving pulse, the vector potential of $E_{\rm d}(\tau)$ approaches a constant $A_{\rm d}(\tau)=-\int_{\tau_{\rm di}}^{\tau}\mathrm{d}\tau'\, E_{\rm d}(\tau') \approx A_{\rm f}$ for $\tau>\tau_{\rm df}$.

The echo-generation process can be divided into the three steps illustrated in Fig.~\ref{fig:echo_bi}(b): (i)~the creation of a photocarrier by the excitation pulse, (ii)~intraband acceleration by the driving pulse, and (iii)~echo emission due to the recombination of the photocarrier wavepacket.
In the following, we evaluate the interband current in Eq.~\eqref{eq:2band_current}, considering these steps one by one.

\textit{(i)~The creation of photocarriers by the excitation pulse.}
After the excitation pulse, the probability amplitudes in Eq.~\eqref{eq:analytical_state} are given by
\begin{gather}
\psi^{\rm cb}_k(\tau_{\rm di}) \approx \mathrm{e}^{-\mathrm{i}\varepsilon^{\rm cb}(k)\tau_{\rm di}} \frac{\mathrm{i} \mu_k}{2}\widetilde{E}_{\rm e}^{\rm env}{\bigl(\varepsilon(k)-\omega_{\rm e}\bigr)} \mathrm{e}^{-\mathrm{i}\theta}, \label{eq:exc_psi_cb} \\
\psi^{\rm vb}_k(\tau_{\rm di}) \approx \mathrm{e}^{-\mathrm{i}\varepsilon^{\rm vb}(k)\tau_{\rm di}}, \label{eq:exc_psi_vb}
\end{gather}
at $\tau = \tau_{\rm di}$ ($\gg \tau_{\rm e}$), up to first order in $\mathcal{H}_{\rm e}$.
Here, we use $\int_{-\infty}^{\tau_{\rm di}} \mathrm{d}\tau'\, E_{\rm e}(\tau') \mathrm{e}^{\mathrm{i}\varepsilon(k)\tau'} \approx \widetilde{E}_{\rm e}(\varepsilon(k)) \approx \widetilde{E}_{\rm e}^{\rm env} (\varepsilon(k)-\omega_{\rm e}) \mathrm{e}^{-\mathrm{i}\theta}/2$, where we have adopted $E_{\rm e}(\tau) = 0$ for $\tau > \tau_{\rm di}$ and have used the rotating-wave approximation in the first and second equalities, respectively, and $\widetilde{E}_{\rm e}^{\rm env}(\omega)$ is the Fourier spectrum of $E_{\rm e}^{\rm env}(\tau)$.

Recalling that $\widetilde{E}_{\rm e}(\omega)$ has a narrow spectral width centered at $\omega=\omega_{\rm e}$, we can expand $\varepsilon(k)$ as $\varepsilon(k) \approx \omega_{\rm e} + v(k_{\rm e}) (k-k_{\rm e})$ for each $k_{\rm e}$ that satisfies $\omega_{\rm e}=\varepsilon(k_{\rm e})$.
In this linear approximation, the Fourier spectrum of the excitation pulse is linearly transcribed into the wavefunction of the electron--hole pair, $\psi_{k}^{\mathrm{cb}} \psi_{k}^{\mathrm{vb}*}$, as illustrated in process (i) of Fig.~\ref{fig:echo_bi}(b).
This process is described by a semiclassical picture in which only the central position of the wavepacket is concerned: electron--hole pairs with energy $\omega_{\rm e}$ are present after the excitation, and the electrons and holes move in opposite directions with relative velocity $v(k_{\rm e})$ [see (i') in Fig.~\ref{fig:echo_bi}(b)].

\textit{(ii)~Intraband acceleration by the driving pulse.}
Once the probability amplitudes at $\tau = \tau_{\rm di}$ are obtained from Eqs.~\eqref{eq:exc_psi_cb} and \eqref{eq:exc_psi_vb}, their time evolution is given exactly by
\begin{align}
\psi^{\lambda}_{k-A_{\rm d}(\tau)}(\tau)=\mathrm{e}^{-\mathrm{i}\int_{\tau_{\rm di}}^{\tau} \mathrm{d}\tau'\, \varepsilon^{\lambda}(k-A_{\rm d}(\tau'))} \psi^{\lambda}_{k}(\tau_{\rm di})
\label{eq:drive_psi}
\end{align}
for any $E_{\rm d}(\tau)$~\cite{Dunlap1986,Korsch2003,Hartmann2004}, where $A_{\rm d}(\tau) =-\int_{\tau_{\rm di}}^{\tau} \mathrm{d}\tau'\, E_{\rm d}(\tau')$ denotes the vector potential of the driving pulse.
The derivation of Eq.~\eqref{eq:drive_psi} is presented in Appendix~\ref{sec:app_bi_echo_analytical}.
The vector potential $A_{\rm d}(\tau)$ of the driving pulse leads to the translation of the electron momentum from $k_{\rm e}$ to $k_{\rm e}-A_{\rm d}(\tau)$, as depicted in process (ii) of Fig.~\ref{fig:echo_bi}(b).
In the semiclassical picture, the relative group velocity of the driven electron--hole pair shifts from $v(k_{\rm e})$ to $v(k_{\rm e}-A_{\rm d}(\tau))$ [see (ii') in Fig.~\ref{fig:echo_bi}(b)].

\textit{(iii)~Echo emission due to the recombination of the photocarrier wavepacket.}
Substituting Eqs.~\eqref{eq:exc_psi_cb}--\eqref{eq:drive_psi} into Eq.~\eqref{eq:2band_current}, we have
\begin{align}
&J_{\mathrm{inter}}(\tau) \notag \\
&\approx \frac{1}{N} \sum_{k} \varepsilon(k-A_{\rm f}) \biggl[ \frac{\mu_{k-A_{\rm f}}^{*} \mu_k}{2} \mathrm{e}^{-\mathrm{i}\varepsilon(k)\tau_{\rm di}} \mathrm{e}^{-\mathrm{i}\int_{\tau_{\rm di}}^{\tau_{\rm df}} \mathrm{d}\tau'\, \varepsilon(k-A_{\rm d}(\tau'))} \notag \\
&\quad \times \mathrm{e}^{-\mathrm{i}\varepsilon(k-A_{\rm f})(\tau-\tau_{\rm df})} \widetilde{E}_{\rm e}^{\rm env}(\varepsilon(k)-\omega_{\rm e}) \mathrm{e}^{-\mathrm{i}\theta}
+ \mathrm{c.c.} \biggr]
\label{eq:acurrent_1}
\end{align}
for $\tau > \tau_{\rm df}$, where we have assumed that the exponent on the right-hand side of Eq.~\eqref{eq:drive_psi} can be rewritten as $\int_{\tau_{\rm di}}^{\tau} \mathrm{d}\tau'\, \varepsilon(k-A_{\rm d}(\tau')) = \int_{\tau_{\rm di}}^{\tau_{\rm df}} \mathrm{d}\tau'\, \varepsilon(k-A_{\rm d}(\tau')) + \varepsilon(k-A_{\rm f})(\tau-\tau_{\rm df})$ with $A_{\rm f} = A_{\rm d}(\tau_{\rm df})$.
Since the spectral width of $\widetilde{E}_{\rm e}^{\rm env}(\omega-\omega_{\rm e})$ is narrow and $\widetilde{E}_{\rm e}^{\rm env}(\omega-\omega_{\rm e})$ is centered at $\omega=\omega_{\rm e}$, the dispersion relations in the rapidly varying functions---i.e., $\widetilde{E}_{\rm e}^{\rm env}$ and the exponentials in Eq.~\eqref{eq:acurrent_1}---can be expanded around $k = k_{\rm e}$, and the summation $\sum_{k}$ can be approximated by the Fourier integral $(2\pi)^{-1} N \int_{-\infty}^{\infty} \mathrm{d}k$.
The other slowly oscillating prefactors are replaced with those for $k=k_{\rm e}$.
Finally, we obtain the expression for the current at $\tau > \tau_{\rm df}$ as
\begin{align}
J_\mathrm{inter}(\tau)
&\approx \sum_{k_{\rm e}} \biggl[ m(k_{\rm e})
E_{\mathrm{e}}^{\mathrm{env}} {\biggl( \frac{v(k_{\rm e}-A_{\rm f})}{v(k_{\rm e})} \tau + \Delta \tau \biggr)} \notag \\
&\quad \times \mathrm{e}^{ -\mathrm{i} [ \varepsilon(k_{\rm e}-A_{\rm f})\tau + \Delta \theta ] } + \mathrm{c.c.} \biggr].
\label{eq:acurrent}
\end{align}
Here, the summation in Eq.~\eqref{eq:acurrent} is over all $k_{\rm e}$ such that $\omega_{\rm e} = \varepsilon(k_{\rm e})$, and $m$, $\Delta\tau$, and $\Delta\theta$ are miscellaneous constants given by
\begingroup\allowdisplaybreaks\begin{gather}
m(k_{\rm e}) = \frac{\varepsilon(k_{\rm e}-A_{\rm f}) }{2v(k_{\rm e})} \mu_{k_{\rm e}-A_{\rm f}}^{*} \mu_{k_{\rm e}}, \\
\Delta \tau = -\frac{v(k_{\rm e}-A_{\rm f})}{v(k_{\rm e})} \tau_{\rm df}
+ \int_{\tau_{\rm di}}^{\tau_{\rm df}} \mathrm{d}\tau'\, \frac{v(k_{\rm e}-A_{\mathrm{d}}(\tau'))}{v(k_{\rm e})} + \tau_{\rm di}, \\
\Delta \theta= \theta -\varepsilon(k_{\rm e}-A_{\rm f}) \tau_{\rm df} + \int_{\tau_{\rm di}}^{\tau_{\rm df}} \mathrm{d}\tau'\, \varepsilon(k_{\rm e}-A_{\mathrm{d}}(\tau')) + \omega_{\rm e} \tau_{\rm di}.
\end{gather}\endgroup
When $v(k_{\rm e}-A_{\rm f})/v(k_{\rm e}) < 0$, as shown in process (iii) of Fig.~\ref{fig:echo_bi}(b), the envelope of the current in Eq.~\eqref{eq:acurrent} is the time-reversed waveform of the envelope of the excitation pulse.
This can be viewed as the reversal of the electron--hole pair distribution in the frequency domain [from the blue to the red curves on the vertical axis in Fig.~\ref{fig:echo_bi}(b)].

In the semiclassical picture, when the relative velocity $v(k_{\rm e}-A_{\rm f})$ is reversed, the electron and hole turn back and eventually recombine, as depicted in process (iii') in Fig.~\ref{fig:echo_bi}(b).
Equation~\eqref{eq:acurrent} then tells us when the echo current is generated.
Since the excitation pulse is centered at $\tau=\tau_{\rm e}$, the echo current is centered at the time when $\tau$ satisfies $[v(k_{\rm e}-A_{\rm f})/v(k_{\rm e})] \tau + \Delta\tau = \tau_{\rm e}$; i.e.,
\begin{align}
\int_{\tau_{\rm e}}^{\tau} \mathrm{d}\tau'\, v(k_{\rm e}-A_{\rm d}(\tau')) = 0. \label{eq:echo_condition}
\end{align}
The condition given in Eq.~\eqref{eq:echo_condition} means that the relative displacement of the photocarriers must be zero when they recombine.
This conclusion is consistent with earlier results obtained by the saddle-point approximation~\cite{Vampa2014,Crosse2014}, although our formula in Eq.~\eqref{eq:acurrent} describes the explicit time dependence of the envelope function beyond the saddle-point approximation.

Another consequence of Eq.~\eqref{eq:acurrent} is that the central frequency of the echo pulse is $\varepsilon(k_{\rm e}-A_{\rm f})$, which depends on the residual vector potential $A_{\rm f}$ of the driving pulse.
This indicates that we are able to obtain the energy-band structure by varying the amplitude of the driving pulse and measuring the frequency of the echo pulses.
In Sec.~\ref{sec:spectroscopy}, we demonstrate numerically that the dispersion relations of quasiparticles can be reconstructed by using the energy-band echoes.

We next make some remarks about the requirement and generality of the energy-band echoes.
First, the frequency of the excitation pulse should not be located at a band edge.
Otherwise, the linear approximation for $\varepsilon(k)$ at $k=k_{\rm e}$ fails to describe the echo current in Eq.~\eqref{eq:acurrent}, even though a remnant of the echoes can still be observed; this is discussed in detail in Appendix~\ref{sec:app_band_edge}.
Second, the present time-reversal dynamics are caused by the reversal of the relative group velocity $v(k_{\rm e}-A_{\rm f})$ of the photoexcited particles, whereas earlier proposals in, e.g., Ref.~\cite{Yuan2016}, require the sign reversal of the total Hamiltonian.
Therefore, for a wide class of insulators in which an electric field can drive well-defined quasiparticles in reciprocal space, energy-band echoes can be observed, and the analytical results in Eqs.~\eqref{eq:acurrent} and \eqref{eq:echo_condition} are valid to some extent.
In these systems, $\varepsilon(k) = \varepsilon^{\rm cb}(k)-\varepsilon^{\rm vb}(k)$ can be thought of as the energy of the photocarriers---e.g., a doublon and holon in Mott insulators---and the relative velocity $v(k)$ in Eq.~\eqref{eq:echo_condition} should be replaced with a vector $\bm{v}(\bm{k})$ in two- or three-dimensional systems.
These conjectures are supported by the numerical calculations presented in Sec.~\ref{sec:spectroscopy}.

%================================================================
%================================================================
\section{Momentum-resolved spectroscopy of quasiparticle excitations} \label{sec:spectroscopy}
In this section, we show that energy-band echoes can be used to reconstruct the dispersion relations of photoexcited quasiparticles.
This is based on the fact that the echo current given in Eq.~\eqref{eq:acurrent} depends on the vector potential $A_{\rm f}$ associated with the amplitude of the driving pulse.
Furthermore, we demonstrate that this scheme can be applied not only to band insulators but also to strongly correlated systems, including Mott and charge-ordered insulators.
We also discuss the effects of integrability and dimensionality on the energy-band echoes.
For simplicity, throughout this section we apply a Gaussian excitation pulse
\begin{align}
A_{\rm e}(\tau) = -A_{\rm e} \mathrm{e}^{-(\tau-\tau_{\rm e})^2/(2\sigma_{\rm e}^2)} \sin (\omega_{\rm e}\tau)
\end{align}
and the subsequent driving pulse given in Eq.~\eqref{eq:drivingpulse}; the parameter values are the same as those adopted in Sec.~\ref{sec:asym_echo} unless otherwise stated.

\subsection{Band insulator} \label{sec:1dbi}
%===============================================
\begin{figure}[t]
\centering
\includegraphics[width=\columnwidth]{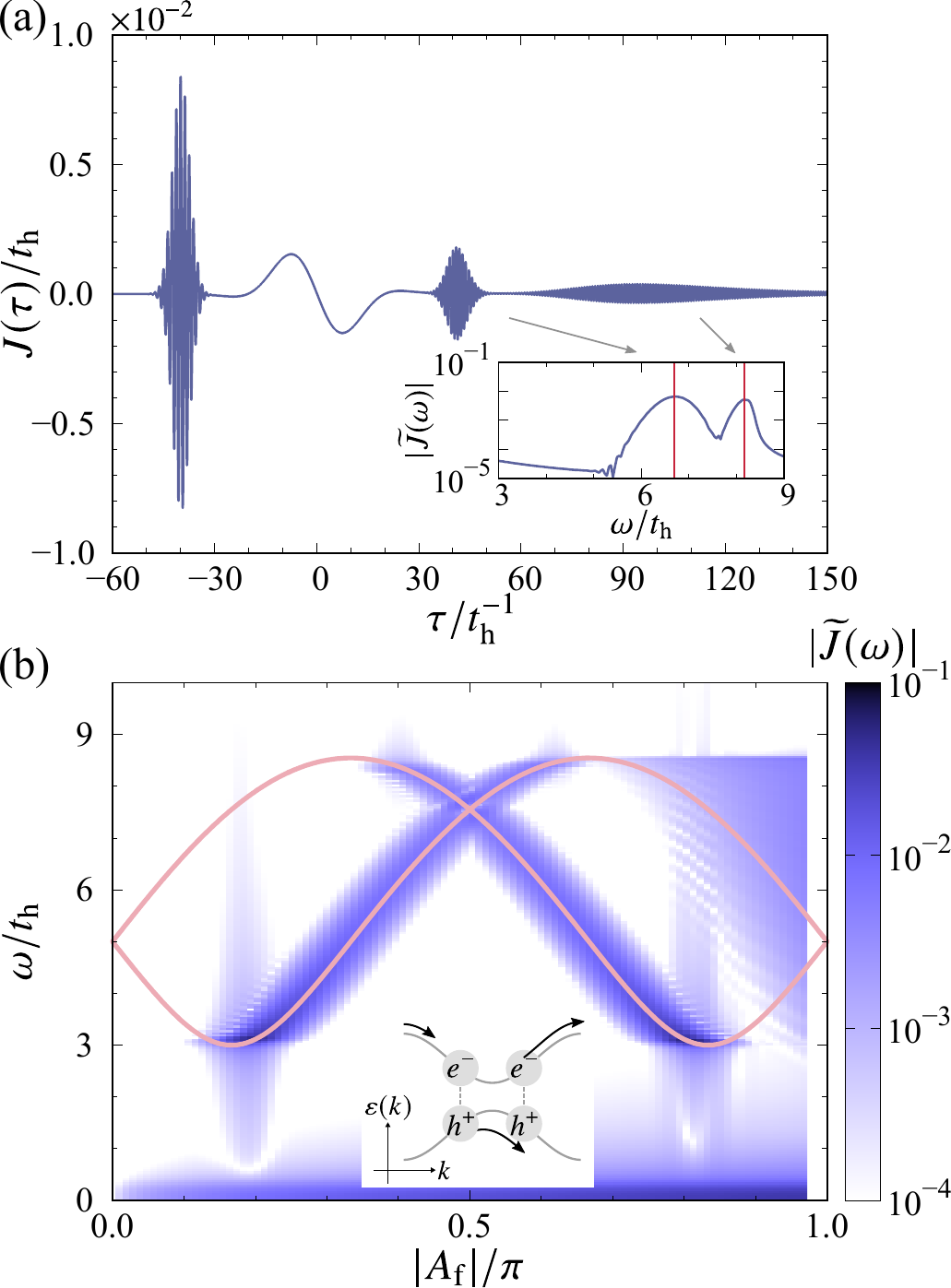}
\caption{Echoes in a one-dimensional band insulator.
(a)~Time profile of the electric current.
The inset shows the Fourier spectrum of the electric current (solid curve) and the energy of the electron--hole pair $\varepsilon_{\rm 1bi}(k_{\rm e}-A_{\rm f})$ calculated from Eq.~\eqref{eq:1bi_disp} (vertical lines).
The amplitude of the driving pulse is set to $A_{\rm d} = 0.9$ (i.e., $A_{\rm f}/\pi \approx -0.44$).
(b)~Spectral map of the electric current.
The red curves show $\omega = \varepsilon_{\rm 1bi}(k_{\rm e}-A_{\rm f})$ as a function of $A_{\rm f}$.
The inset shows a sketch of the intraband dynamics of two photoexcited pairs.
}
\label{fig:echo_1dbi}
\end{figure}
%===============================================

In Fig.~\ref{fig:echo_1dbi}, we show the energy-band echoes that appear in the one-dimensional band insulator defined by Eq.~\eqref{eq:biH}.
After the driving pulse decays, two echo pulses appear at $\tau \approx 40t_{\rm h}^{-1}$ and $\tau \approx 90t_{\rm h}^{-1}$.
The appearance of the two pulses can be ascribed to the fact that the equation $\omega_{\rm e} = \varepsilon(k_{\rm e})$ has two roots in the one-dimensional system.
The roots of $\omega_{\rm e} = \varepsilon(k_{\rm e})$ in the first Brillouin zone $(-\pi, \pi]$ are $k_{\rm e} \approx \pm 1.05$ and $\pm 2.1$; the energies of the electron--hole pairs after the driving pulse are given by $\varepsilon_{\rm 1bi}(k_{\rm e}-A_{\rm f}) \approx 6.69t_{\rm h}$ for $k_{\rm e} \approx +1.05$ and $-2.1$, and $8.16t_{\rm h}$ for $k_{\rm e} \approx -1.05$ and $+2.1$.
These values are in good agreement with the peak structure of the Fourier spectrum of the echo current for $\tau \geq 0$, denoted by $\widetilde{J}(\omega)$, as shown in the inset of Fig.~\ref{fig:echo_1dbi}(a).
The frequency of the first (second) echo at $\tau \approx 40 t_{\rm h}^{-1}$ ($90 t_{\rm h}^{-1}$) is found to be $\omega \approx 6.69 t_{\rm h}$ ($8.16 t_{\rm h}$).
The first echo was already seen in Fig.~\ref{fig:echo_bi}(a) for the excitation pulse defined by Eq.~\eqref{eq:excitationpulse_sec2}.

We next consider the $A_{\rm f}$ dependence of the echo frequency in more detail.
Figure~\ref{fig:echo_1dbi}(b) shows a color map of the Fourier spectra of $J(\tau)$ for $\tau > 0$.
There are two branches: with increasing $|A_{\rm f}|$, one increases from $\omega = 3t_{\rm h}$ to $8.54t_{\rm h}$ and the other decreases from $\omega = 8.54t_{\rm h}$ to $3t_{\rm h}$.
These branches are described quite well by the equation $\omega = \varepsilon_{\rm 1bi}(k_{\rm e}-A_{\rm f})$ as indicated by the red curves in Fig.~\ref{fig:echo_1dbi}(b).
We also find that the echo appears for $A_{\rm f}$ between the first and second extrema of $\omega = \varepsilon_{\rm 1bi}(k_{\rm e}-A_{\rm f})$, where the relative velocity of the electron--hole pair is reversed by the driving pulse, as illustrated in the inset of Fig.~\ref{fig:echo_1dbi}(b).
These results are consistent with the analysis in Sec.~\ref{sec:analysis} and demonstrate that the frequency of the echo provides information about the energy-band structure $\varepsilon_{\rm 1bi}(k_{\rm e}-A_{\rm f})$ as a function of the vector potential $A_{\rm f}$.

When the excitation pulse is resonant with a band edge, however, the analytical expressions derived in Sec.~\ref{sec:analysis} are not valid because $v(k_{\rm e}) = 0$ in Eq.~\eqref{eq:acurrent}.
In this case, the frequency of the echoes deviates from $\omega = \varepsilon_{\mathrm{1bi}}(k_\mathrm{e}-A_\mathrm{f})$, as discussed in Appendix~\ref{sec:app_band_edge}.

\subsection{Mott insulator} \label{sec:mott}
Mott insulators comprise another important class of insulators, in which the electron--electron interaction energy dominates over the kinetic energy of an electron.
Here, we consider a prototypical model of a Mott insulator, i.e., the half-filled one-dimensional Hubbard model.
The Hamiltonian for this system is given by
\begin{align}
\mathcal{H}_{\rm H} = -\sum_{is} {\left( t_{\rm h} c_{is}^{\dagger} c_{i+1,s} +{\rm H.c.} \right)} +U\sum_{i} n_{i\uparrow} n_{i\downarrow},
\label{eq:hubH}
\end{align}
where $c_{is}^{\dagger}$ ($c_{is}$) is the creation (annihilation) operator for an electron at site $i$ with spin $s\ (={\uparrow},{\downarrow})$, and $n_{i}=\sum_s n_{is}=\sum_s c_{is}^{\dagger} c_{is}$ represents the number operator.
The nearest-neighbor transfer integral and the on-site repulsive interaction are denoted by $t_{\rm h}$ and $U$, respectively.
The ground state at half filling is a Mott insulating state, and the elementary charge excitations are doublons and holons.
According to the Bethe ansatz~\cite{Essler2005}, the energy and momentum of a doublon--holon pair can be written as $\mathcal{E}(k_{\rm d}, k_{\rm h}) = \varepsilon_{\rm d}(k_{\rm d}) + \varepsilon_{\rm h}(k_{\rm h})$ and $P(k_{\rm d}, k_{\rm h}) = p_{\rm d}(k_{\rm d}) + p_{\rm h}(k_{\rm h})$, respectively, where
\begingroup\allowdisplaybreaks\begin{align}
\varepsilon_{\rm h}(k) &= \varepsilon_{\rm d}(k)  \notag \\
&= 2t_{\rm h}\cos k +\frac{U}{2} \notag \\ 
&\ \ \ \ + 2\int_0^{\infty} \frac{\mathrm{d}\omega}{\omega}\frac{\mathcal{J}_1(\omega)\cos(\omega\sin k) \mathrm{e}^{-\omega U/(4t_{\rm h})} }{\cosh[\omega U/(4t_{\rm h})]},
\label{eq:doublon_holon_energy} \\
p_{\rm h}(k) &= p_{\rm d}(k)+\pi \notag \\
&= \frac{\pi}{2}-k-2\int_0^{\infty} \frac{\mathrm{d}\omega}{\omega}\frac{\mathcal{J}_0(\omega)\sin (\omega\sin k )}{1+\mathrm{e}^{\omega U/(2t_{\rm h})}}.
\label{eq:doublon_holon_momentum}
\end{align}\endgroup
Here, $k_{\rm d}$ ($k_{\rm h}$) is called the spectral parameter for a doublon (holon), and $\mathcal{J}_n$ denotes the $n$th-order Bessel function of the first kind.
Since we consider optical excitations with $P(k_{\rm d}, k_{\rm h}) = 0$, the spectral parameters satisfy the relation $k_{\rm d} = -k_{\rm h}$.
The energy of the photoinduced doublon--holon pair is given by $\varepsilon_{\rm H}(k) = \mathcal{E}(k,-k)$.
The spectral parameter in the photoexcited state, denoted by $k=k_{\rm e}$, is given by the roots of the equation $\omega_{\rm e} = \varepsilon_{\rm H}(k_{\rm e})$, and the parameter $k$ after the driving pulse is determined by $k = p^{-1}(p(k_{\rm e})-A_{\rm f})$, with $p$ being the shifted doublon momentum $p(k) = p_{\rm d}(k) + \pi/2 \ (= p_{\rm h}(k) - \pi/2)$.

We obtain the ground state and simulate the real-time dynamics numerically by using the infinite density-matrix renormalization group (iDMRG) and the infinite time-evolving block decimation (iTEBD) methods, respectively~\cite{mcculloch2008infinite,Vidal2007}, for matrix product states with global $U(1) \otimes U(1)$ symmetry associated with the conservation of the total number and magnetization of the electrons~\cite{Hastings2009}.
The bond dimension of the matrix product states is set to $\chi = 400$ for the ground states and $\chi = 2000$ for the time-evolved states; the relative error of the ground-state energy is of order $10^{-6}$.
The vector potential $A(\tau)$ is introduced as the Peierls phase; i.e., $t_{\rm h} \to t_{\rm h} \mathrm{e}^{-\mathrm{i}A(\tau)}$.
The real-time evolution of $|\psi(\tau)\rangle$ is given by $|\psi(\tau+\delta\tau)\rangle \approx \exp[-\mathrm{i}\mathcal{H}_{\rm H}(\tau+\delta \tau/2) \delta \tau] \vert \psi(\tau) \rangle$, and the second-order Suzuki--Trotter decomposition is adopted.

%===============================================
\begin{figure}[t]
\centering
\includegraphics[width=\columnwidth]{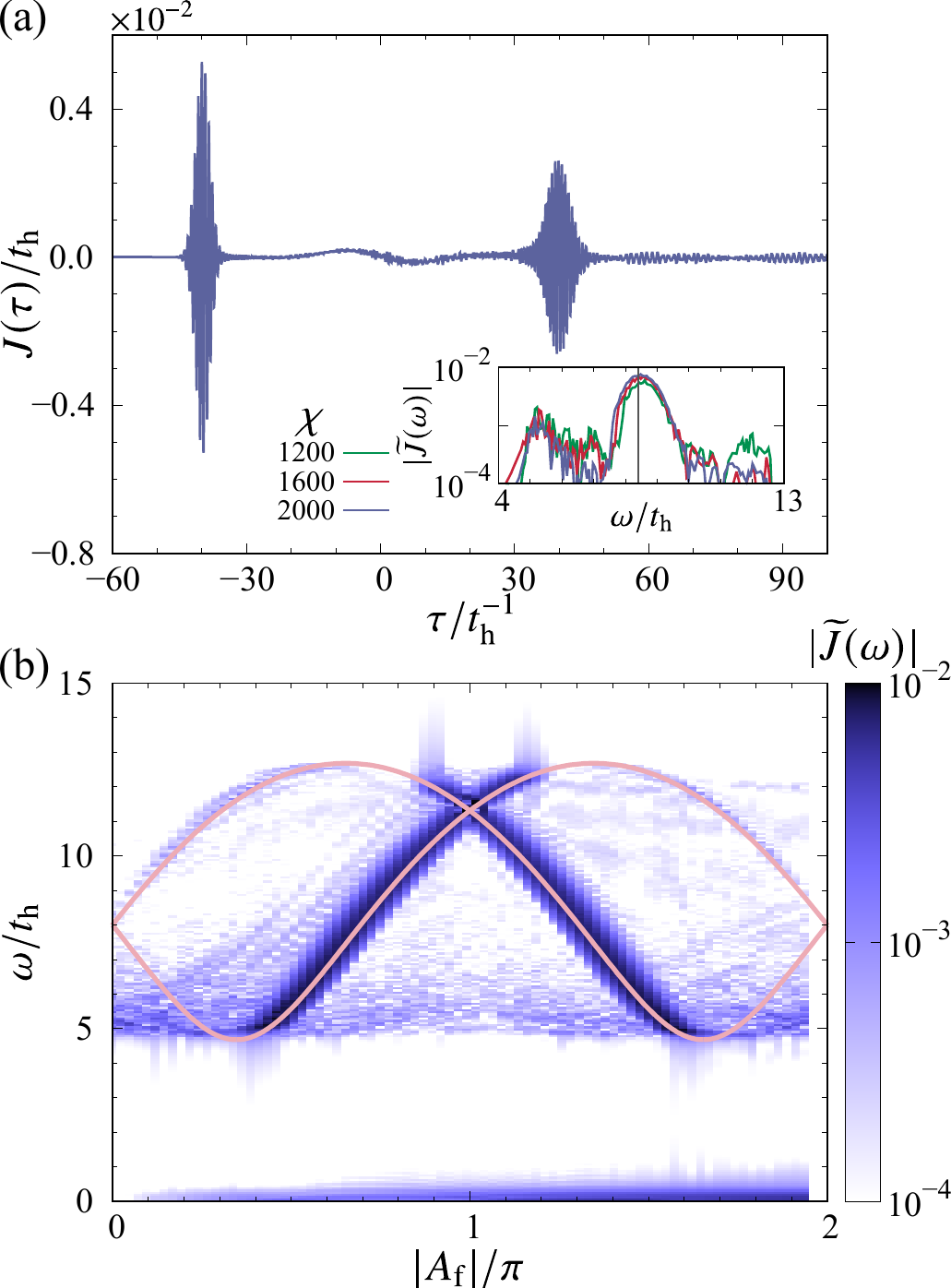}
\caption{Echoes in a Mott insulator.
(a)~Time profile of the electric current calculated by the iTEBD method.
The amplitude of the driving pulse is set to $A_{\rm d} = 1.5$ (i.e., $A_{\rm f}/\pi \approx -0.73$) and the other parameters are $\omega_{\rm e} = U = 8 t_{\rm h}$ and $A_{\rm e} = 0.004$.
The inset shows the Fourier spectra for different bond dimensions $\chi$ and the energy of the doublon--holon pair.
(b)~Spectral map of the electric current.
The bond dimension is set to $\chi = 2000$.
The red curves show $\omega = \varepsilon_{\rm H}(p^{-1}(p(k_{\rm e})-A_{\rm f}))$ as a function of $A_{\rm f}$.
}
\label{fig:echo_mott}
\end{figure}
%===============================================

Figure~\ref{fig:echo_mott}(a) shows the calculated time profile for the electric current in a one-dimensional Mott insulator with $U = 8t_{\rm h}$.
The frequency of the excitation pulse is set to $\omega_{\rm e}=8t_{\rm h}$; the corresponding spectral parameter is $k_{\rm e} \approx \pm 1.73$.
We find that the echo current is induced at $\tau \approx 40 t_{\rm h}^{-1}$.
The inset in Fig.~\ref{fig:echo_mott}(a) shows that the spectral peak of the echo is located at $\omega \approx 8.41 t_{\rm h}$ in consistent with the energy calculated from $\varepsilon_{\rm H}(p^{-1}(p(k_{\rm e})-A_{\rm f}))$ and denoted by the vertical line; we also observe convergence with respect to $\chi$.

The Fourier spectra of the echoes are displayed in Fig.~\ref{fig:echo_mott}(b).
There are two clear branches, which are similar to those in the band insulator and which are in agreement with the exact doublon--holon energies denoted by the red curves.
This observation shows that energy-band echoes can be used to obtain the quasiparticle dispersion relations even in a strongly correlated insulator, whereas one-particle spectra observed by angle-resolved photoemission spectroscopy are smeared in such a system since non-interacting electrons are not the well-defined quasiparticles any longer~\cite{Kim2006,Benthien2007}.

\subsection{Charge-ordered insulator} \label{sec:co}
A strong interaction often favors long-range order that spontaneously breaks the symmetry of a system.
In this section, we consider a charge-ordered insulator without inversion symmetry as an example to gain further insight into the energy-band echoes.

We adopt the one-dimensional transverse-field Ising (TFI) model.
The Hamiltonian for this system is given by
\begin{align}
\mathcal{H}_{\rm TFI} = - V \sum_i \sigma_i^z \sigma_{i+1}^z -t_{\rm h}\sum_i \sigma_i^{x}, \label{eq:tfiH}
\end{align}
where $\{\sigma_i^x, \sigma_i^y, \sigma_i^z\}$ are the Pauli matrices at the $i$th unit cell.
The first and second terms in Eq.~\eqref{eq:tfiH} represent the Ising interaction and the transverse field, respectively.
By using the Jordan--Wigner transformation, Pfeuty~\cite{Pfeuty1970} rigorously showed that the ground state is a spontaneously-broken-symmetry state with $\sum_i \langle \sigma_i^z \rangle \neq 0$ for $t_{\rm h} < V$, and it undergoes a phase transition to a disordered phase with $\sum_i \langle \sigma_i^z \rangle = 0$ via the quantum critical point at $t_{\rm h} = V$.
Using the Fourier and Bogoliubov transformations, we can reduce the Hamiltonian to the diagonal form
\begin{align}
\mathcal{H}_{\rm TFI} &= \sum_{k} \varepsilon_{1}(k) \eta_k^\dagger \eta_k, \\
\varepsilon_1(k) &= 2\sqrt{t_{\rm h}^2 + V^2 + 2t_{\rm h}V \cos k }, \label{eq:tfi_1_disp}
\end{align}
where $\eta_k^\dagger$ denotes a fermionic creation operator with momentum $k$.
In the ordered phase, the elementary excitation is a kink or domain-wall excitation.
The photoexcited state is given by $\eta_{k}^\dagger \eta_{-k}^\dagger |0\rangle$~\cite{Imai2019q}, with $|0\rangle$ being the vacuum (i.e., the ground state), in which a kink--antikink pair has the energy
\begin{align}
\varepsilon_{\rm TFI}(k) = \varepsilon_1(k) + \varepsilon_1(-k). \label{eq:tfi_disp}
\end{align}

%===============================================
\begin{figure}[t]
\centering
\includegraphics[width=\columnwidth]{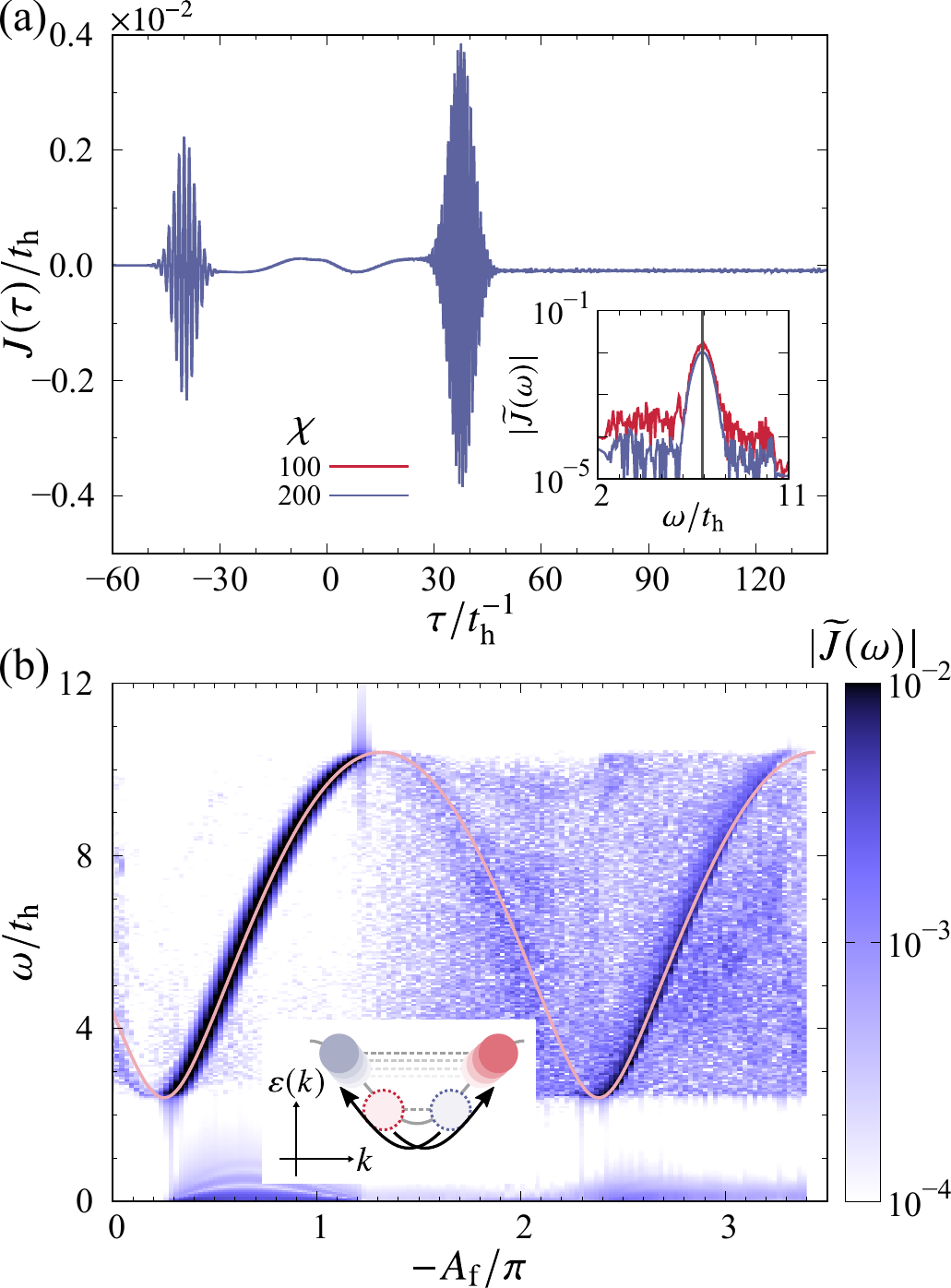}
\caption{Echoes in a charge-ordered insulator.
(a)~Time profile of the electric current calculated by the iTEBD method.
The amplitude of the driving pulse is set to $A_{\rm d} = 1.5$ (i.e., $A_{\rm f}/\pi \approx -0.73$), and the other parameters are $\omega_{\rm e} = 4.4 t_{\rm h}$ and $V = 1.6 t_{\rm h}$.
The inset shows the Fourier spectra for $\chi = 100$ and $200$; the energy of the kink--antikink pair is indicated by the vertical line.
(b)~Spectral map of the electric current.
The red curve shows $\omega = \varepsilon_{\rm TFI}(k_{\rm e} - P_0 A_{\rm f})$ as a function of $A_{\rm f}$ with $P_0 \approx 0.94$ (see text).
The inset in (b) illustrates the motion of the kink--antikink pair in reciprocal space.
}
\label{fig:echo_ti}
\end{figure}
%===============================================

Notwithstanding its simplicity, the TFI model emerges in various contexts in physics.
For example, it is known that organic ferroelectrics that consist of molecular dimers can be described by the TFI model~\cite{Naka2010,Hotta2010,Imai2019q}, with the Pauli matrices representing the intradimer orbital degree of freedom, $t_{\rm h}$ denoting the intradimer transfer integral, and $V$ being proportional to the interdimer repulsive interaction strength.
The broken-symmetry phase is interpreted as a charge-ordered phase with a finite electric polarization $\sum_i \langle \sigma_i^z \rangle \neq 0$.
The Peierls substitution of the vector potential $A$ yields a rotation of the transverse field around the $z$-axis; i.e.,
\begin{align}
\mathcal{H}_{\rm TFI}(A) = - V\sum_i \sigma_i^z \sigma_{i+1}^z - t_{\rm h}\sum_i {\left[ \sigma_i^{x} \cos A - \sigma_i^{y} \sin A \right]}.
\label{eq:tfiH_A}
\end{align}
By introducing the time-dependent unitary transformation $\mathcal{U} = \exp[-\mathrm{i}A(\tau) \sum_i \sigma_i^z/2]$~\cite{Imai2019q}, we can rewrite Eq.~\eqref{eq:tfiH_A} as
\begin{align}
\mathcal{H}_{\rm TFI}(E) = - V \sum_i \sigma_i^z \sigma_{i+1}^z -t_{\rm h}\sum_i \sigma_i^{x} - E(\tau) \sum_{i} \frac{\sigma_i^z}{2},
\end{align}
with $E(\tau) = -\partial_\tau A(\tau)$.
The polarization operator is defined by
\begin{align}
\hat{P} = - \frac{1}{N} \frac{\delta\mathcal{H}_{\rm TFI}}{\delta E} = \frac{1}{2N} \sum_{i} \sigma_i^z.
\end{align}
We obtain the ground state of the Hamiltonian given in Eq.~\eqref{eq:tfiH} and simulate the real-time evolution governed by the Hamiltonian in Eq.~\eqref{eq:tfiH_A} by using the iTEBD method with $\chi = 200$; the absolute error of the ground-state energy is of the order of $10^{-8} t_{\rm h}$.
Here, we chose a positively polarized state with $\langle 0 | \hat{P} | 0 \rangle > 0$ as the initial state.

Figure~\ref{fig:echo_ti}(a) displays the time profile of the electric current and its Fourier spectrum in the charge-ordered phase with $V = 1.6 t_{\rm h}$.
The excitation-pulse frequency is set to $\omega_{\rm e} = 4.4 t_{\rm h}$, which creates a kink--antikink pair with $k = k_{\rm e} \approx 2.396$.
An echo pulse is generated at $\tau \approx 40 t_{\rm h}^{-1}$, and its central frequency converges with respect to $\chi$, as shown in the inset.
We show the spectral map of the electric current in Fig.~\ref{fig:echo_ti}(b).
The equation $\omega_{\rm e} = \varepsilon_{\rm TFI}(k)$ has the single root $k = k_{\rm e}$, and there is a single branch given by $\omega = \varepsilon_{\rm TFI}(k_{\rm e} - P_0 A_{\rm f})$~[the vertical line in the inset of Fig.~\ref{fig:echo_ti}(a) and the red curve in Fig.~\ref{fig:echo_ti}(b)].
Here, the prefactor of $A_{\rm f}$ is defined by $P_0 = 2 \langle 0 | \hat{P} | 0 \rangle$, which originates from a non-local operator appearing in the Jordan--Wigner transformation and reflects the absence of inversion symmetry; see Appendix~\ref{sec:app_drive_tfi} for a detailed discussion.
In this sense, energy-band echoes capture the many-body nature associated with spontaneous symmetry breaking.

%===============================================
\begin{figure*}[t]
\centering
\includegraphics[width=0.92\hsize]{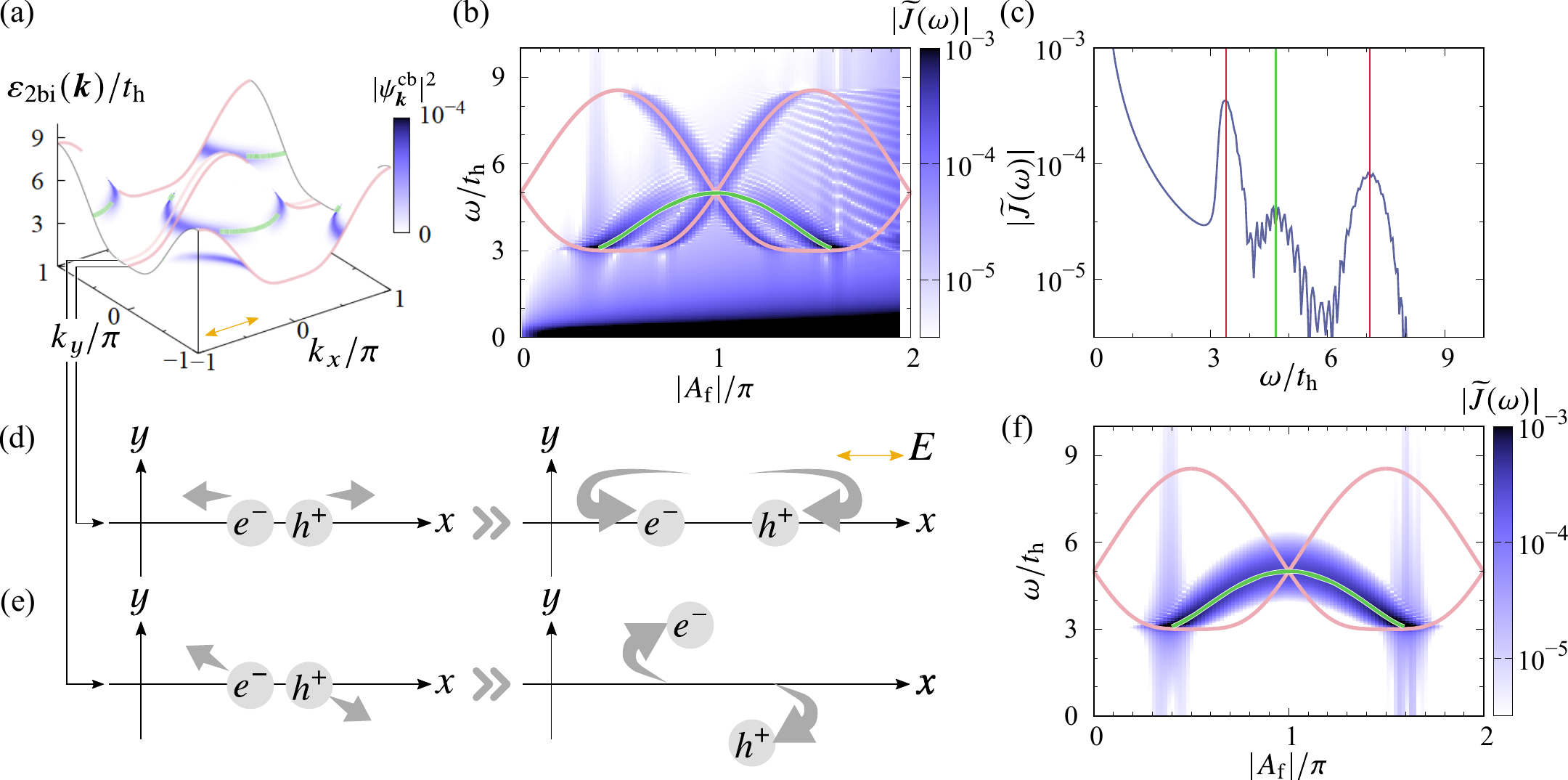}
\caption{Echoes in a two-dimensional band insulator.
(a)~Energy-band structure $\varepsilon_{\rm 2bi}(\bm{k})$.
The contrast on the surface represents $|\psi_{\bm{k}}^{\mathrm{cb}}|^2$ for an $x$-polarized excitation pulse.
The double-headed arrow indicates the polarization of the excitation pulse.
(b)~Spectral map of the electric current with an $x$-polarized excitation pulse and an $x$-polarized driving pulse.
The red and green curves represent $\omega = \varepsilon_{\rm 2bi}(k_{\mathrm{e}, x}^{(1)}-A_{\rm f},0)$ and $\varepsilon_{\rm 2bi}(k_{\mathrm{e}, x}^{(2)}-A_{\rm f},\pi)$, and $\omega = \varepsilon_{\rm 2bi}(k_{{\rm e},x}-A_{\rm f},k_{{\rm e},y})$, respectively (see text).
(c)~Fourier spectrum of the electric current for $A_{\rm d} = 1.64$ (i.e., $A_{\rm f}/\pi \approx -0.79$).
The vertical lines represent the calculated energies of the quasiparticles after the driving pulse.
(d),(e)~Sketches of the motions of the photoexcited electron and hole under an $x$-polarized driving pulse.
(f)~Spectral map of the electric current with a $y$-polarized excitation pulse and an $x$-polarized driving pulse.
The parameters are set to $\omega_{\rm d}=0.1t_{\rm h}$, $\omega_{\rm e}=5t_{\rm h}$, $A_{\rm e}=0.002$, and $L=1500$ in (a)--(f).
}
\label{fig:echo_2dbi}
\end{figure*}
%===============================================

\subsection{Dimensionality and integrability}
\label{sec:dimensionality}
We have hitherto discussed energy-band echoes with a focus on one-dimensional integrable systems, in which the photoexcited quasiparticles have an infinite lifetime.
A natural question then arises: whether and how do energy-band echoes appear in higher-dimensional or non-integrable systems?
In the following, we address this question by considering a two-dimensional tight-binding model and a one-dimensional extended Hubbard model.

\subsubsection{Two-dimensional band insulator} \label{sec:2dbi}
We adopt the two-orbital tight-binding Hamiltonian given in Eq.~\eqref{eq:biH} on a square lattice with nearest-neighbor hopping.
The energy of an electron--hole pair is given by
\begin{align}
\varepsilon_{\rm 2bi}(\bm{k}) = 2\sqrt{[2t^{\alpha\beta}(\cos k_x + \cos k_y )]^2+(E_{\rm g}/2)^2},
\label{eq:2bi_disp}
\end{align}
where $\bm{k} = (k_x, k_y)$ is the momentum of the electron.
We simulate the real-time dynamics as in Sec.~\ref{sec:asym_echo}, with the insulating initial state having $N/L^2 = 1$, where $L^2$ denotes the number of sites.
The parameter values are $E_{\rm g} = 3t_{\rm h}$, $t^{\alpha\beta} = t_{\rm h}$, and $\omega_{\rm e} = 5 t_{\rm h}$.

First, we apply the excitation pulse in the $x$-direction.
Figure~\ref{fig:echo_2dbi}(a) shows the quasiparticle band structure $\varepsilon_{\rm 2bi}(\bm{k})$ and the conduction-electron distribution $|\psi_{\bm{k}}^{\mathrm{cb}}|^2$ after the excitation pulse.
The electrons are excited to the iso-energy surface given by $\omega_{\rm e} = \varepsilon_{\rm 2bi}(\bm{k})$.
Since the transition moment is proportional to the relative group velocity $\bm{v}_{2{\rm bi}}(\bm{k}) = \partial \varepsilon_{\rm 2bi}(\bm{k})/\partial \bm{k}$, the electron distribution $|\psi_{\bm{k}}^{\mathrm{cb}}|^2$ vanishes where $v_{\mathrm{2bi},x}(\bm{k})=0$.

When we apply the driving pulse in the $x$-direction, we observe echo generation even in this two-dimensional system, and we obtain the spectral map of the electric current shown in Fig.~\ref{fig:echo_2dbi}(b).
This panel contains three branches: two of them, ranging from $\omega = 3 t_{\rm h}$ to $8.544 t_{\rm h}$, can be attributed to quasiparticles excited at $\bm{k} = (k_{\mathrm{e}, x}^{(1)}, 0)$ with $\omega_{\rm e} = \varepsilon_{\rm 2bi}(k_{\mathrm{e}, x}^{(1)}, 0)$ and at $\bm{k} = (k_{\mathrm{e}, x}^{(2)}, \pi)$ with $\omega_{\rm e} = \varepsilon_{\rm 2bi}(k_{\mathrm{e}, x}^{(2)}, \pi)$.
These are in agreement with $\omega = \varepsilon_{\rm 2bi}(k_{\mathrm{e}, x}^{(1)} - A_{\rm f}, 0)$ and $\omega = \varepsilon_{\rm 2bi}(k_{\mathrm{e}, x}^{(2)} - A_{\rm f}, \pi)$, as indicated by the red curves in Fig.~\ref{fig:echo_2dbi}(b) and the red vertical lines in Fig.~\ref{fig:echo_2dbi}(c).
Since $v_{\mathrm{2bi},y}(k_x,0) = 0$ for any $k_x$, these quasiparticles move in the $x$-direction parallel to the driving pulse, as depicted in Fig.~\ref{fig:echo_2dbi}(d).
For quasiparticles with momenta that depart slightly from $k_{y}=0$, the electron and hole cannot recombine, since the sign of $v_{\mathrm{2bi},y}(\bm{k})$ is unchanged by the driving pulse, as illustrated in Fig.~\ref{fig:echo_2dbi}(e), and thus they do not generate echoes.
These two branches can therefore be understood by analogy to those in the one-dimensional systems.

The third branch in Fig.~\ref{fig:echo_2dbi}(b) stems from the two-dimensional motions of quasiparticles excited with nonzero relative velocity in the $y$-direction, and it is reproduced by the following analysis of the classical motion of the quasiparticles.
Assuming that Eq.~\eqref{eq:echo_condition} holds for higher-dimensional systems, we expect an echo to appear when the relative displacement $\bm{r}(\tau)$ is zero after the driving pulse.
This condition is given by
\begin{align}
\bm{r}(\tau) = \int_{\tau_{\rm e}}^{\tau} \mathrm{d}\tau'\, \bm{v}_{\rm 2bi}(\bm{k}_{\rm e}-\bm{A}(\tau')) = 0, \label{eq:echo_condition_vector}
\end{align}
where $\bm{k}_{\rm e}$ satisfies $\omega_{\rm e} = \varepsilon_{\rm 2bi}(\bm{k}_{\rm e})$.
From Eq.~\eqref{eq:echo_condition_vector}, we obtain a set of $\bm{k}_{\rm e}$ that contributes to the echoes [indicated by the green dots in Fig.~\ref{fig:echo_2dbi}(a)] and the corresponding echo frequency $\omega = \varepsilon_{\rm 2bi}(k_{{\rm e},x}-A_{\rm f},k_{{\rm e},y})$ [the green curve in Fig.~\ref{fig:echo_2dbi}(b) and the green vertical line in Fig.~\ref{fig:echo_2dbi}(c)].

By changing the polarization of the excitation pulse, we can distinguish these two types of echoes mentioned above: one comes from one-dimensional motions and the other from higher-dimensional motions.
Figure~\ref{fig:echo_2dbi}(f) shows a spectral map of the electric current with a $y$-polarized excitation pulse and an $x$-polarized driving pulse; we observe only the single branch attributed to the two-dimensional motions, although the echo intensity is different from that in Fig.~\ref{fig:echo_2dbi}(b).
Since quasiparticles with $k_y = 0$ are not excited by the $y$-polarized pulse in the present model, the echoes originating from the one-dimensional motion completely disappear.
Therefore, in principle, we can reconstruct the energy-band structure of two- or three-dimensional materials through a comprehensive analysis of the polarization dependence of the energy-band echoes.

\subsubsection{Non-integrable Mott insulator} \label{sec:uv}
Next, we consider the one-dimensional extended Hubbard model defined by the Hamiltonian
\begin{align}
\mathcal{H}_{\rm extH} = \mathcal{H}_{\rm H} + V\sum_{i}n_{i}n_{i+1},
\label{eq:ehubH}
\end{align}
where the first term $\mathcal{H}_{\rm H}$ is the Hubbard Hamiltonian given in Eq.~\eqref{eq:hubH}, and the second term represents a repulsive interaction between the nearest-neighbor sites.
The integrability of the system is broken when $V \neq 0$, which is supported by the observation that the level-spacing statistics are close to the Wigner--Dyson distribution, as shown in Appendix~\ref{sec:app_level_statistics}.

\begin{figure}[t]
\centering
\includegraphics[width=\columnwidth]{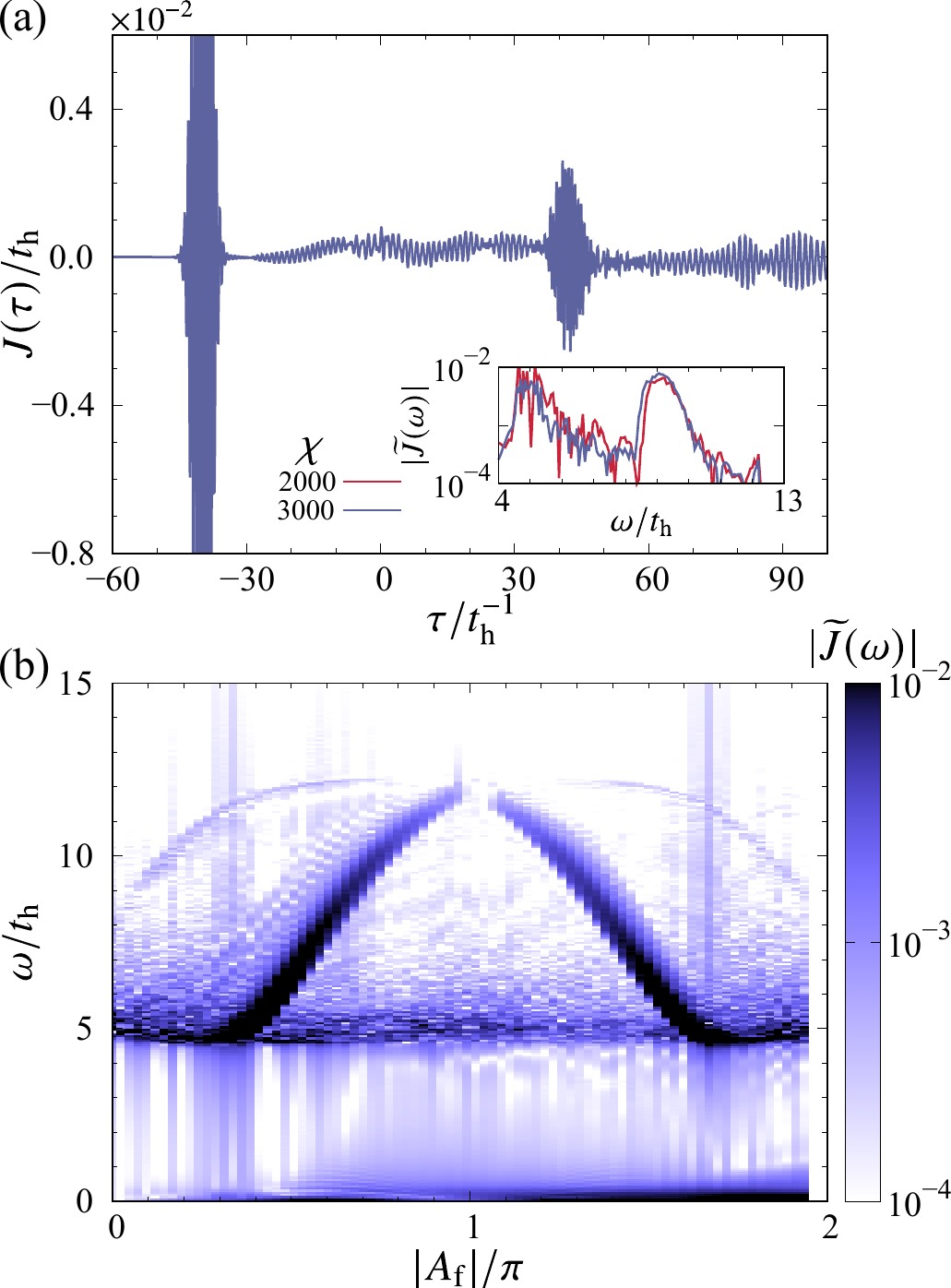}
\caption{Echoes in a non-integrable Mott insulator.
(a)~Time profile of the electric current calculated by the iTEBD method.
The amplitude of the driving pulse is set to $A_{\rm d} = 1.4$ (i.e., $A_{\rm f}/\pi \approx -0.68$), and the other parameters are $\omega_{\rm e} = U = 8t_{\rm h}$, $V = 2t_{\rm h}$, and $A_{\rm e} = 0.02$.
The inset shows the Fourier spectra for $\chi = 2000$ and $3000$.
(b)~Spectral map of the electric current.
The bond dimension is set to $\chi = 3000$.
}
\label{fig:echo_uv}
\end{figure}

By using the iDMRG and iTEBD methods as in Sec.~\ref{sec:mott}, we can calculate the ground state in the Mott insulating phase ($U = 8t_{\rm h}$ and $V = 2t_{\rm h}$) with $\chi = 400$ and its real-time evolution with $\chi = 3000$.
The time profile of the electric current is shown in Fig.~\ref{fig:echo_uv}(a), and its Fourier spectrum is plotted in the inset.
The echo signal is generated at $\tau \approx 40 t_{\rm h}^{-1}$, whose spectral peak is located at $\omega \approx 9t_{\rm h}$.
A small oscillation with frequency $\omega \approx 5t_{\rm h}$ is also observed after the excitation pulse decays, which we attribute to the excitation of bound doublon--holon pairs, i.e., excitons~\cite{Jeckelmann2003}.
The bound pair does not give rise to echoes since it is not driven by the electric field.

Figure~\ref{fig:echo_uv}(b) shows a spectral map of the electric current.
Prominent spectral weight emerges at $|A_{\rm f}|/\pi \approx 0.3$, transfers from $\omega \approx 5t_{\rm h}$ to $12t_{\rm h}$ with increasing $|A_{\rm f}|$, and then vanishes at $|A_{\rm f}|/\pi = 1$; it appears again for $|A_{\rm f}|/\pi > 1$ and transfers downward.
Although no exact expression for the quasiparticle energy is known, these two prominent peaks are likely to belong to two different branches that are adiabatically connected to those of the Hubbard model with $V=0$, providing information about the quasiparticle dispersion relations as a function of $A_{\rm f}$.
We attribute the non-dispersive peak at $\omega \approx 5t_{\rm h}$ to the excitation of bound pairs.

As shown above, energy-band echoes can be observed even in a non-integrable system, in which the photoexcited quasiparticles acquire a finite lifetime.
Considering the similarity between the energy-band echoes and Hahn echoes, we anticipate that the relaxation or dephasing time of the quasiparticles can be estimated from the delay-time dependence of the echo intensity.
However, we leave this issue for future work, since the present numerical method does not accurately capture the long-time behavior.

%===============================================
%===============================================

\subsection{Optical driving pulse} \label{sec:optical}
%===============================================
\begin{figure}[t]
\centering
\includegraphics[width=\columnwidth]{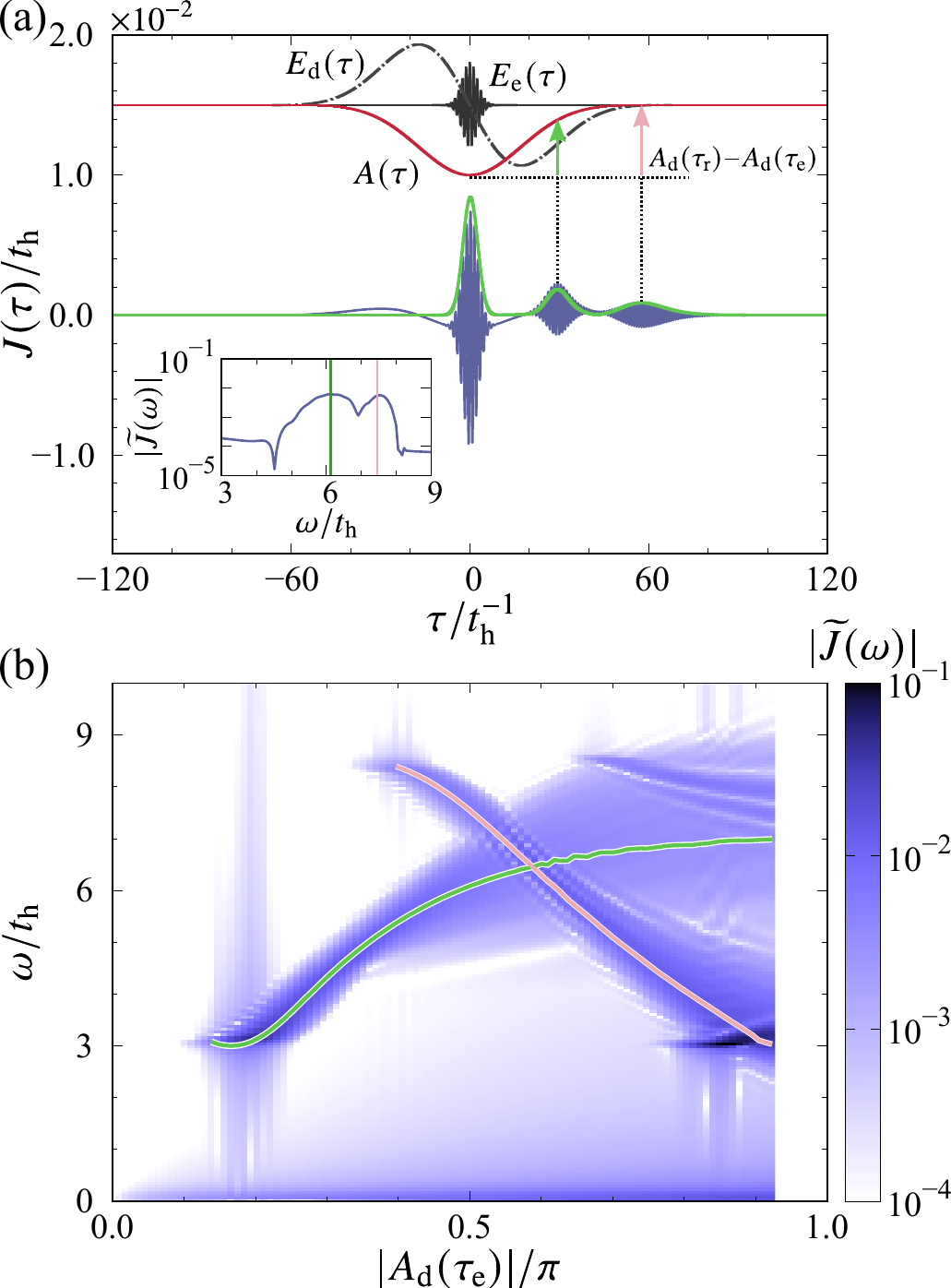}
\caption{Echoes in a one-dimensional band insulator with an optical driving pulse.
(a)~Time profiles of the electric fields, vector potential, and electric current.
The green curve shows the envelope of $J(\tau)$ calculated from the Hilbert transformation.
The amplitude of the driving pulse is set to $A_{\rm d} = 2.2$ (i.e., $|A_{\mathrm{d}}(\tau_{\mathrm{e}})|/\pi \approx 0.51$), and the other parameters are $\omega_{\rm d} = 0.05t_{\rm h}$, $\omega_{\rm e} = 5t_{\rm h}$, $A_{\rm e} = 0.002$, $\tau_{\rm e}=0$, and $N = 1500$.
The inset shows the Fourier spectrum of $J(\tau)$.
(b)~Spectral map of the electric current.
The green and red curves represent $\omega = \varepsilon_{\rm 1bi}(k_{\rm e} - A_{\mathrm{d}}(\tau_{\mathrm{r}}) + A_{\mathrm{d}}(\tau_{\mathrm{e}}))$ as functions of $|A_{\mathrm{d}}(\tau_{\mathrm{e}})|$ for the first and second echoes, respectively, where $A_{\mathrm{d}}(\tau_{\mathrm{r}}) - A_{\mathrm{d}}(\tau_{\mathrm{e}})$ is the vector-potential change to which the electron--hole pairs are subjected.
}
\label{fig:optical_setup}
\end{figure}
%===============================================

In previous sections, we used the driving pulse $E_{\rm d}(\tau)$ given in Eq.~\eqref{eq:drivingpulse}, with the CEP being $\theta_{\rm d} = 0$, which leaves a finite vector potential $A_{\rm f} \propto \cos \theta_{\rm d}$ after $E_{\rm d}(\tau)$ decays.
While the existence of such a unipolar (i.e., $A_{\rm f} \neq 0$) pulse is controversial~\cite{Bessonov1981,Kim2000b,Arkhipov2020,Arkhipov2022}, here by introducing a CEP-controlled mono-cycle driving pulse we show that energy-band echoes can be obtained even for $A_{\rm f} = 0$.

We adopt a driving pulse $E_{\rm d}(\tau)$ with CEP $\theta_{\rm d} = \pi/2$ that contains a single optical cycle, and we apply the excitation pulse at $\tau = \tau_{\mathrm{e}} = 0$ when the vector potential of the driving pulse is maximal.
The excitation pulse must be shorter than a half-cycle of the driving pulse, which means that the pulse width $\sigma_{\rm e}$ must satisfy $\omega_{\rm e}^{-1} \ll \sigma_{\rm e} \ll \omega_{\rm d}^{-1}$.
The waveforms of $E_{\rm d}(\tau)$, $E_{\rm e}(\tau)$, and $A(\tau) = A_{\rm d}(\tau)+A_{\rm e}(\tau)$ are shown in Fig.~\ref{fig:optical_setup}(a).
With this setup, the photoexcited quasiparticles are subjected to a finite impulse due to the driving pulse even though $A_{\rm f} = 0$.
The impulse is determined by the vector potential at $\tau = \tau_{\mathrm{e}}$, denoted by $A_{\mathrm{d}}(\tau_{\mathrm{e}})$ [see Fig.~\ref{fig:optical_setup}(a)].

Figure~\ref{fig:optical_setup}(a) also shows the time profile of the electric current and its spectrum for the one-dimensional band insulator defined by $\mathcal{H}_{\mathrm{bi}}$ in Eq.~\eqref{eq:biH}.
We find that two echoes appear at $\tau \approx 30t_{\rm h}^{-1}$ and $60t_{\rm h}^{-1}$.
We show the spectral map of the electric current in Fig.~\ref{fig:optical_setup}(b), where two branches are seen, although they are broadened for large $|A_{\mathrm{d}}(\tau_{\mathrm{e}})|$ compared with the previous results.

To compare the energy-band dispersion with the echo frequency, we evaluate the changes in the vector potential between $\tau=\tau_{\rm e}$ and the time $\tau = \tau_{\rm r}$ at which the quasiparticles recombine and the echoes are generated.
The green curves in Fig.~\ref{fig:optical_setup}(a) represent the envelope of $J(\tau)$ obtained through a low-pass filter and the Hilbert transformation, from which we evaluate the changes in $A(\tau)$ for the first and second echoes.
The vertical lines in the inset of Fig.~\ref{fig:optical_setup}(a) and the green and red curves in Fig.~\ref{fig:optical_setup}(b) show $\omega = \varepsilon_{\rm 1bi}(k_{\rm e}-A_{\mathrm{d}}(\tau_{\rm r})+A_{\mathrm{d}}(\tau_{\mathrm{e}}))$, which agree with the simulated spectra of the echoes.

%================================================================
%================================================================
\section{Discussion} \label{sec:discussion}
We have illustrated the appearance of energy-band echoes in band insulators and correlated insulators.
We find that what is essential for the generation of the energy-band echoes is the presence of an energy continuum of photoexcited quasiparticles that can be driven by the external electric field.
Therefore, we can expect to observe such echoes not only in solids but also in a cold-atom system on an optical lattice.
In such a system, resonant interband excitation has been achieved~\cite{Fischer1998}, and an intraband driving has been realized by using a constant inertial force~\cite{BenDahan1996} and a synthetic gauge field~\cite{Lin2011,Pineiro2019}.

For energy-band echoes in solids, dissipation of the photocarriers may not be negligible, which reduces the echo intensity, as we mentioned in Sec.~\ref{sec:uv} above.
To observe these echoes, one optical cycle of the driving pulse is preferably shorter than the relaxation times of the photocarriers, which are typically of the order of a few femtoseconds for electron--electron interactions and subpicoseconds for electron--phonon interactions~\cite{Kruchinin2019a}.
Hence, to prevent relaxation during the driving process, a femtosecond excitation pulse and a mono-cycle mid-infrared or terahertz driving pulse are suitable for experiments.
In the present study, we used $\omega_{\rm d} = 0.1t_{\rm h} \sim 0.1\ \mathrm{eV}$ (i.e., $\hbar/\omega_{\rm d} \sim 6.6\ \mathrm{fs}$), which is much shorter than the typical timescales of electron--phonon interactions but longer than those of electron--electron interactions.
Nevertheless, echoes are still observed in a non-integrable Mott insulator with $U = 4V = 8t_{\rm h} \sim 8\ \mathrm{eV} \gg \omega_{\rm d}$, as shown in Sec.~\ref{sec:uv}, which leaves open the possibility of observing such echoes in real materials.

The energy-band echoes require both an excitation pulse and a driving pulse, which is reminiscent of high-order sideband generation (HSG)~\cite{Kono1998,Zaks2012,Langer2016,Banks2017,Langer2018,Uchida2018,Borsch2020,Nagai2020a}.
In the HSG process, an excited state is prepared by a resonant pulse and then driven by an intense multicycle terahertz pulse.
In the present study, however, the vector-potential shift due to the half- or mono-cycle driving pulse uncovers such a light-emission process during a single optical cycle, which leads to the concept of energy-band echoes.

Angle-resolved photoemission spectroscopy (ARPES) is a well-established and sophisticated experimental method that directly accesses the energy bands below the Fermi level.
Since the present spectroscopy based on energy-band echoes (energy-band echo spectroscopy for short) acquires the dispersion relations of the electron--hole pairs, we can obtain the band structure below and above the Fermi level in combination with ARPES, which is similar to an earlier proposal for all-optical spectroscopy based on HHG~\cite{Vampa2015b}.
Furthermore, energy-band echo spectroscopy provides well-defined dispersion relations for renormalized quasiparticles such as the doublon--holon pairs in a Mott insulator, as shown in Sec.~\ref{sec:mott}, whereas ARPES spectra for this case are usually blurred because of many-body interactions~\cite{Kim2006,Benthien2007}.

%================================================================
%================================================================
\section{Summary} \label{sec:summary}
In this work, we have investigated energy-band echoes that originate from the dynamics of quasiparticle wavepackets driven and controlled by a lightwave.
After the driving pulse decays, the electric current oscillates with the time-reversed waveform of the excitation pulse.
The echoes are observed not only in band insulators but also in correlated insulators, in one and higher dimensions.
On the basis of the numerical and analytical results, we have elucidated the echo-generation process: (i)~a photocarrier is excited, and its wavepacket has the same waveform as the excitation pulse; (ii)~the photocarrier is adiabatically accelerated by the driving pulse; and (iii)~the recombination of the wavepacket yields echoes with the time-reversed waveform.
Furthermore, we found that the dispersion relation is reflected in the echo frequency as a function of the driving-pulse amplitude.
We have confirmed numerically that the echo frequency agrees with the predictions obtained from exact solutions, and we have also found that the echoes appear even in a non-integrable system where the quasiparticles have a finite lifetime.
These results suggest that energy-band echoes can be used to achieve momentum-resolved spectroscopy of elementary optical excitations in a wide class of insulators.

\begin{acknowledgments}
One of the authors, Sumio Ishihara, passed away in November 2020 during the preparation of the manuscript.
The authors thank Yusuke~Masaki for fruitful discussions and Joji~Nasu for useful comments on the manuscript.
This work was supported by JSPS KAKENHI, Grant Nos.\ JP21J10575, JP18H05208, JP19K23419, and JP20K14394.
Some of the numerical calculations were performed using the facilities of the Supercomputer Center, the Institute for Solid State Physics, The University of Tokyo.
\end{acknowledgments}

%================================================================
%================================================================
\appendix
\section{Intraband dynamics in a band insulator}
\label{sec:app_bi_echo_analytical}
We derive Eq.~\eqref{eq:drive_psi} by considering the electric-field-induced dynamics of a non-interacting tight-binding model~\cite{Dunlap1986,Korsch2003,Hartmann2004}.
The Hamiltonian for orbital $\lambda$ can be written as
\begin{align}
\mathcal{H}_{\lambda} = \sum_{m=0}^{\infty} {\left[ g_m^{\lambda}{\hat{K}}^m + g_m^{\lambda *} (\hat{K}^{\dagger})^m \right]} -E_{\rm d}(\tau) \hat{R},
\label{eq:driveH}
\end{align}
where $\hat{K}^m=\sum_j c_{\lambda,j}^{\dagger} c_{\lambda,j+m}$ and $\hat{R}=\sum_j r_{j} c_{\lambda,j}^{\dagger} c_{\lambda,j}$, and $g_{m}^{\lambda}$ is defined by $\varepsilon^{\lambda}(k)=  \sum_{m=0}^{\infty} \mathrm{e}^{\mathrm{i}km}g_m^{\lambda} + \mathrm{c.c}$.
We introduce the time-evolution operator $U(\tau) = U_R(\tau) U_K(\tau)$ as
\begin{align}
\mathrm{i} \partial_{\tau} U_{R}(\tau) &= -E_{\rm d}(\tau) \hat{R} U_{R},
\label{eq:difeq_U_N} \\
\mathrm{i} \partial_{\tau} U_{K}(\tau) &= U_{R}^{-1} \sum_{m} {\left[ g_m^{\lambda} \hat{K}^m + g_m^{\lambda *} (\hat{K}^{\dagger})^m \right]} U_{R} U_{K}.
\label{eq:difeq_U_K}
\end{align}
The solution of Eq.~\eqref{eq:difeq_U_N} is given by
\begin{align}
U_{R}(\tau) &= \mathrm{e}^{-\mathrm{i} A_{\rm d}(\tau) \hat{R}},
\label{eq:root_U_N}
\end{align}
where $A_{\rm d}(\tau)=-\int_{\tau_{\rm di}}^{\tau} \mathrm{d}\tau'\, E_{\rm d}(\tau')$ is the vector potential of the driving pulse.
Since $[\hat{R},\hat{K}^m] = -m\hat{K}^m$ and $[\hat{K},\hat{K}^{\dagger}] = 0$ in the thermodynamic limit, we obtain the solution of Eq.~\eqref{eq:difeq_U_K} in the form
\begin{align}
U_{K}(\tau) = \exp {\left\{ {-\mathrm{i} \sum_m {\left[ \chi_m^{\lambda}(\tau) \hat{K}^m + \chi_m^{\lambda}(\tau)^* (\hat{K}^{\dagger})^m \right]}} \right\} } ,
\label{eq:root_U_K}
\end{align}
where $\chi_m^{\lambda}(\tau) = \int_{\tau_{\rm di}}^{\tau} \mathrm{d}\tau'\, g_m^{\lambda} \mathrm{e}^{-\mathrm{i}A_{\rm d}(\tau') m}$.
The matrix elements of $U(\tau)$ are given by
\begin{align}
\langle k | U(\tau) | k'\rangle = \delta_{k+A_{\rm d}(\tau),k'}\, \mathrm{e}^{-\mathrm{i} \int_{\tau_{\rm di}}^{\tau} \mathrm{d}\tau'\, \varepsilon^{\lambda}(k'-A_{\rm d}(\tau'))}
\label{eq:element_U}
\end{align}
with $|k\rangle = N^{-1/2}\sum_{j} \mathrm{e}^{\mathrm{i}kr_j} c_{\lambda,j}^\dagger |0\rangle$.
Substituting Eq.~\eqref{eq:element_U} into $\psi_k(\tau)=\sum_{k'} \langle k| U(\tau) | k'\rangle \psi_{k'}(\tau_{\rm di})$, we obtain Eq.~\eqref{eq:drive_psi}.

\section{Band-edge excitation}
\label{sec:app_band_edge}
%===============================================
\begin{figure}[t]
\centering
\includegraphics[width=\columnwidth]{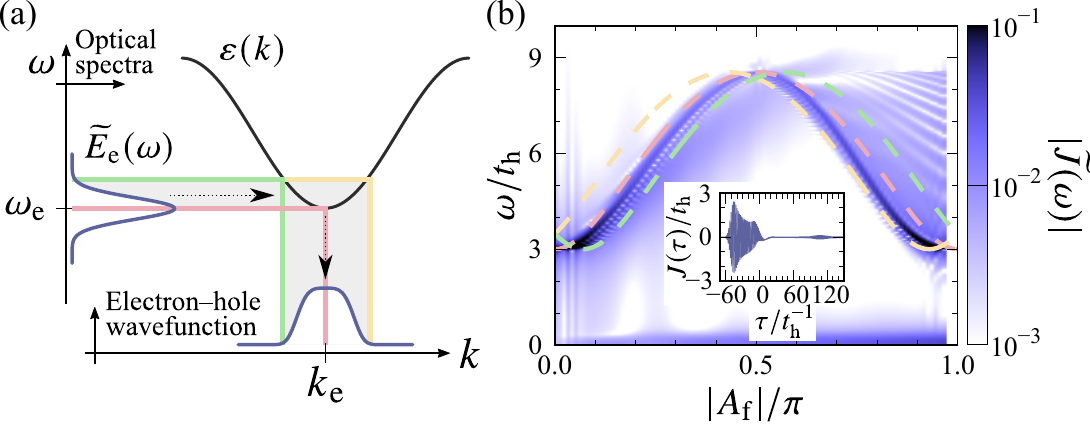}
\caption{Light emission for a band-edge excitation pulse.
(a)~Sketch of band-edge excitation.
Only a part of $\widetilde{E}(\omega)$ (gray shading) can be transferred to the wavefunction of the electron--hole pair.
(b)~Spectral map of the electric current.
The red, yellow, and green dashed curves represent $\omega = \varepsilon_{\mathrm{1bi}}(k_{\mathrm{e}}-A_{\mathrm{f}})$, $\varepsilon_{\mathrm{1bi}}(k_{\mathrm{e}} + \Delta k -A_{\mathrm{f}})$, and $\varepsilon_{\mathrm{1bi}}(k_{\mathrm{e}} - \Delta k -A_{\mathrm{f}})$, respectively, with $k_\mathrm{e} = \pi/2$ and $\Delta k \approx 0.23$.
The inset shows the time profile of the electric current for $A_{\rm d}=1.08$ (i.e., $A_{\rm f}/\pi \approx -0.52$).
The parameters are set to $\omega_{\rm d}=0.1t_{\rm h}$, $A_{\rm e}=0.002$, $\tau_{\rm e}=-50t_{\rm h}^{-1}$, and $\omega_{\rm e}=3t_{\rm h}=E_{\rm g}$.
}
\label{fig:edge_echo}
\end{figure}
%===============================================

The expression for the electric current given in Eq.~\eqref{eq:acurrent} does not hold when the excitation pulse is resonant with a band edge.
In this case, only a part of the Fourier spectrum of the excitation pulse is transcribed into the electron--hole wavefunction, as shown in Fig.~\ref{fig:edge_echo}(a).
This gives rise to the collapse of the time-reversed waveform for $\tau > 0$.

In Fig.~\ref{fig:edge_echo}(b), we show the time profile and spectral map of the electric current in a one-dimensional band insulator for a band-edge excitation pulse with $\omega_\mathrm{e} = E_\mathrm{g}$; the other parameters are the same as in Sec.~\ref{sec:1dbi}.
The oscillation from time $\tau \approx -40t_{\mathrm{h}}^{-1}$ to $-20t_{\mathrm{h}}$ is due to the presence of electron--hole pairs with zero group velocity.
A remnant of the echoes is observed in $J(\tau)$ at $\tau \approx 100t_\mathrm{h}$, as shown in the inset of Fig.~\ref{fig:edge_echo}(b).
The Fourier spectra of $J(\tau)$, shown in Fig.~\ref{fig:edge_echo}(b), indicate the appearance of a dispersive peak, although it deviates from $\omega = \varepsilon_{\mathrm{1bi}}(k_{\mathrm{e}} - A_{\mathrm{f}})$ with $k_\mathrm{e} = \pi/2$.
Considering the spectral width of the excitation pulse, depicted in Fig.~\ref{fig:edge_echo}(a), we also plot two curves, $\omega = \varepsilon_{\mathrm{1bi}}(k_\mathrm{e} + \Delta k - A_{\mathrm{f}})$ and $\varepsilon_{\mathrm{1bi}}(k_\mathrm{e} - \Delta k - A_{\mathrm{f}})$, where $\Delta k$ satisfies $\varepsilon_{\mathrm{1bi}}(k_\mathrm{e} \pm \Delta k) = E_\mathrm{g} + 2\sigma_{\mathrm{e}}^{-1}$, in Fig.~\ref{fig:edge_echo}(b).
The peak frequency is found to lie between the two curves.

\section{Intraband dynamics of the one-dimensional transverse-field Ising model}
\label{sec:app_drive_tfi}
Here, we explain how the correction factor $P_0$ emerges in the energy-band echoes in the one-dimensional TFI model.
The Jordan--Wigner transformation is defined by
\begin{align}
\sigma_j^x &= 2n_j - 1, \\
S_j^{+} &= c_j^{\dagger} \mathrm{e}^{\mathrm{i}\pi \sum_{l<j} n_l},
\end{align}
where $c_j^{\dagger}$ denotes a fermion creation operator, $n_j=c_j^{\dagger} c_j$ represents the number operator, and $S_j^{\pm} = (\sigma_j^y \pm \mathrm{i} \sigma_j^z)/2$ is a spin raising/lowering operator with the $x$-axis being the quantization axis.
Since the Peierls substitution of the vector potential rotates the transverse field around the $z$-axis, the Hamiltonian in Eq.~\eqref{eq:tfiH_A} can be diagonalized by using the following transformation:
\begin{align}
\widetilde{\sigma}_j^x &= \sigma_j^x \cos A - \sigma_j^y \sin A, \label{eq:sigma_x_tilde} \\
\widetilde{\sigma}_j^y &= \sigma_j^x \sin A + \sigma_j^y \cos A, \label{eq:sigma_y_tilde} \\
\widetilde{\sigma}_j^z &= \sigma_j^z, \label{eq:sigma_z_tilde}
\end{align}
which leaves the eigenvalues of $\mathcal{H}_{\mathrm{TFI}}$ unchanged.
Accordingly, the fermion operator $c_{j}^\dagger$ becomes
\begin{align}
\widetilde{c}_j &= {\left( c_j \frac{\mathrm{e}^{-\mathrm{i}A\sigma_j^z} +1}{2} + c_j^{\dagger} \frac{\mathrm{e}^{-\mathrm{i}A\sigma_j^z}-1}{2} \right)} \mathrm{e}^{-\mathrm{i}A\sum_{l<j} \sigma_l^z}. 
\label{eq:c_dagger_tilde-}
\end{align}
Since what is needed is a one-to-one correspondence of the energy eigenstates between systems with infinitesimally small differences in $A$, by assuming $(\mathrm{e}^{\pm \mathrm{i}A\sigma_j^z} +1)/2 \approx 1$ and $(\mathrm{e}^{\pm \mathrm{i} A \sigma_j^z} -1)/2 \approx 0$ for $|A| \ll 1$, we have
\begin{align}
\widetilde{c}_j \approx c_j \mathrm{e}^{-\mathrm{i}A\sum_{l<j} \sigma_l^z}.
\label{eq:c_dagger_tilde+_approx_sigma_z}
\end{align}
In the energy-band echo process, only the broken-symmetry ground state $|0\rangle$ and low-lying excited states are involved in the dynamics.
In these states, the polarization density is expected to be spatially uniform, and it is given approximately by $P_0 = N^{-1} \sum_{l} \langle 0 | \sigma_l^z | 0 \rangle = \langle 0 | \sigma_l^z | 0 \rangle$ for the excited states as well as for the ground state.
With the additional assumption that quantum fluctuations due to the transverse field are negligible~\footnote{We confirmed numerically that this assumption is valid for $t_{\rm h}/V \lesssim 0.7$ in the TFI model.}, the operator $\sum_{l < j} \sigma_l^z$ in Eq.~\eqref{eq:c_dagger_tilde+_approx_sigma_z} can be replaced with a c-number: $\sum_{l < j} \sigma_l^z \to j P_0$ for a domain wall propagated in the positive direction.
Then we obtain
\begin{align}
\widetilde{\eta}_{-k}^{\dagger} \approx  u_k c_{-(k+P_0 A)}^{\dagger} + \mathrm{i}v_k c_{k+P_0 A},
\label{eq:eta_tilde}
\end{align}
where $c_{k}^{\dagger} = N^{-1/2}\sum_{j} \mathrm{e}^{\mathrm{i}kj}c_j^{\dagger}$, $u_k = \cos(\theta_k/2)$, and $v_k=-\sin(\theta_k/2)$, with $\theta_k = \tan^{-1} [(-V \sin k)/(-t_{\rm h}-V\cos k)]$, which diagonalizes the Hamiltonian.
Equation~\eqref{eq:eta_tilde} indicates that the energy after the driving pulse is given by $\varepsilon_1(k - P_0 A_{\rm f})$.

\section{Level statistics for the extended Hubbard model}
\label{sec:app_level_statistics}
Here, we show numerically the Wigner--Dyson level statistics of the extended Hubbard model on a one-dimensional chain with $L$ sites [Eq.~\eqref{eq:ehubH}].
Since the Hamiltonian has global symmetries---i.e., the translational ($\mathcal{T}$), parity ($\mathcal{P}$), time-reversal ($\varTheta$), particle--hole ($\mathcal{C}$), and spin-rotational symmetries---we consider a subspace that contains a ground state with electron density $N/L = 1$, total magnetization $S^z = 0$, spin quantum number $S = 0$, momentum $k = 0$, and $(\mathcal{P}, \varTheta, \mathcal{C}) = (+1, +1, +1)$ for $L = 0 \bmod 4$ and $(+1, -1, -1)$ for $L = 2 \bmod 4$.
We use the Lanczos method with desymmetrization of the Hamiltonian~\cite{Poilblanc1993,Kudo2005a,Sandvik2010a} and projection onto the subspace with $S = 0$.

%===============================================
\begin{figure}[t]
\centering
\includegraphics[width=\columnwidth]{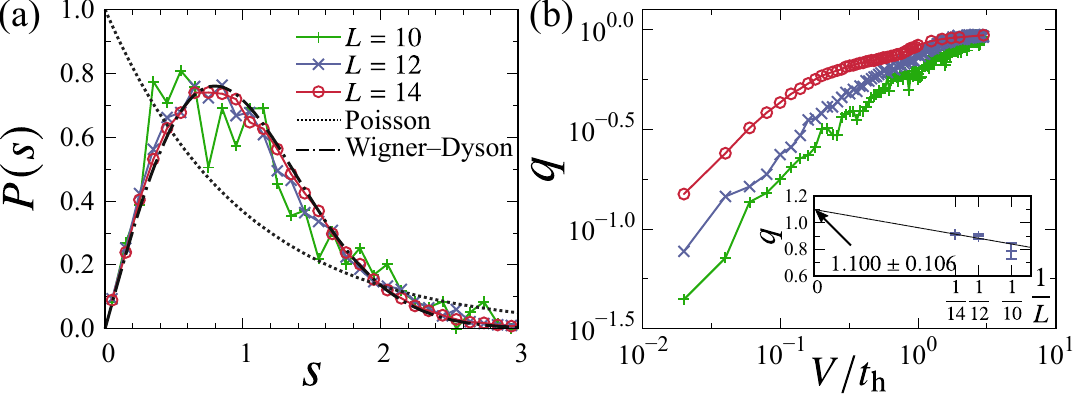}
\caption{Level statistics for the one-dimensional extended Hubbard model.
(a)~Nearset-neighbor level-spacing distribution for $V=2t_{\rm h}$.
The smoothing parameter and bin width are set to $\sigma = 0.5$ and $\delta s = 0.1$, respectively.
(b)~The parameter $q$ as a function of $V$, for $L = 10$--$14$.
The inset shows $q$ for $V=2t_{\rm h}$; the line is obtained by the weighted least-squares method.
}
\label{fig:rmt}
\end{figure}
%===============================================

Figure~\ref{fig:rmt}(a) shows the nearest-neighbor level-spacing distributions $P(s)$ for $L = 10$, $12$, and $14$, where we adopt an unfolding method with the Gaussian kernel density estimation.
With increasing $L$, the distribution $P(s)$ approaches the Wigner--Dyson distribution $P_{\mathrm{WD}}(s) = (\pi/2) s \exp(-\pi^2 s^2/4)$ rather than the Poisson distribution $P_{\mathrm{P}}(s) = \exp(-s)$, indicating that the system is non-integrable when $V = 2t_{\mathrm{h}}$.

To discuss the $V$ dependence of the level-spacing distribution, we introduce the Brody function defined by $P_q(s) = \alpha(q) s^q \exp[-\beta(q) s^{1+q}]$, where $\alpha(q) = (1+q) \beta(q)$ and $\beta(q) = \{ \Gamma[(2+q)/(1+q)] \}^{1+q}$ with $\Gamma$ being the gamma function.
Since $P_q(s)$ reduces to the Poisson distribution for $q = 0$ and to the Wigner--Dyson distribution for $q = 1$, the parameter $q$ interpolates between integrable and non-integrable systems.
We obtain $q$ by fitting $P(s)$ to the Brody function, as shown in Fig.~\ref{fig:rmt}(b).
The parameter $q$ increases with $V$ and approaches $1$ as $L \to \infty$ for $V > 0$.
Therefore, the nearest-neighbor interaction $V$ breaks the integrability, as one naively expects.

\bibliography{reference}
\end{document}